\documentclass[12pt]{JHEP3}
\newcommand{\be}{\begin{equation}}
\newcommand{\ee}{\end{equation}}
\newcommand{\beq}{\begin{equation}}
\newcommand{\eeq}{\end{equation}}
\newcommand{\ba}{\begin{eqnarray}}
\newcommand{\ea}{\end{eqnarray}}
\newcommand{\nn}{\nonumber}

\def\td{{\tilde{d}}}
\def\cT{{\cal T}}

\def\tgamma{\tilde{\gamma}}
\def\roughly#1{\mathrel{\raise.3ex\hbox{$#1$\kern-.95em
\lower1ex\hbox{$\sim$}}}}
\def\lsim{\roughly<}

\def\pref#1{(\ref{#1})}

\def\ssubsubsection#1{\vspace{3mm} \noindent \textbf{#1} \\ \vspace{-3mm} \\ \noindent}

\usepackage{epsfig}

\title{On Bouncing Brane-Worlds, S-branes and
Branonium Cosmology}

\author{C.P. Burgess,$^{1}$ F. Quevedo,$^{2}$ R. Rabad\'an,$^{3}$ G.
Tasinato $^{4}$ and I. Zavala $^{5}$
\\

$^1$ Physics Department, McGill University,
                3600 University Street,\\
                Montr\'eal, Qu\'ebec, Canada, H3A 2T8.\\

$^2$ Centre for Mathematical Sciences, DAMTP,
               University of Cambridge,\\
               Cambridge CB3 0WA UK.\\

$^3$ Theory Division, CERN, Geneva 23, Switzerland.\\

$^4$ Physikalisches Institut der Universit\"at Bonn, \\
    Nussallee 12, 53115 Bonn, Germany.\\

$^5$ Physics Department, University of Colorado,
 Boulder, CO 80309 USA.}

\abstract{We present several higher-dimensional spacetimes for
which observers living on 3-branes experience an induced metric
which bounces. 
The classes of examples include boundary branes on 
generalised 
S-brane backgrounds and probe branes in D-brane/anti D-brane systems.
The bounces we consider normally would be expected
to require an energy density which violates the weak energy
condition, and for our co-dimension one examples this is
attributable  to bulk curvature terms in the
effective Friedmann equation.
We examine the features of the acceleration which provides the
bounce,
 including in some cases the existence of positive acceleration
without event  horizons, and we give a geometrical interpretation for it.
 We discuss the stability of the solutions from the
point of view of both the brane and the bulk. Some of our examples
appear to be stable from the bulk point of view, suggesting the
possible existence of stable bouncing cosmologies within the
brane-world framework.}

 \preprint{McGill-03/19, 
CERN-TH/2003-203, COLO-HEP-493, DAMTP-2003-104}

\begin{document}

\newcommand{\sfrac}[2]{{\textstyle\frac{#1}{#2}}}
\def\eq{Eq.\,}
\def\eqs{Eqs.\,}
\def\be{\begin{equation}}
\def\ee{\end{equation}}
\def\bea{\begin{eqnarray}}
\def\eea{\end{eqnarray}}
\def\M{\mathcal M}
\def\noi{\noindent}
\def\nn{\nonumber}
\def\g{{\rm g}}


\section{Introduction}
Bouncing cosmologies have been advocated as having played a role
in our past, both within pre-Big Bang cosmologies \cite{prebb} 
for which new
string-motivated physics smoothes out the Big Bang singularity, 
and
within cyclic scenarios where the universe survives the passage
through a succession of earlier singularities \cite{ekpirosis}.
 Interest in these
proposals has been further sharpened by the recent precise
measurements of temperature fluctuations in the cosmic microwave
background (CMB). In particular, these models provide the main
alternatives to the inflationary description of these
fluctuations, motivating the detailed study of the kinds of
late-time cosmologies to which they give rise.

A major obstacle to understanding their predictions arises from
their potential dependence on the details of the bounce, and a
study of this dependence has been hindered by the absence of a
well-behaved model of a bouncing universe with which to test
theoretical proposals. The difficulty with making such a model
hinges on the necessity of violating the weak energy condition in
order to do so, since this appears to inevitably require a
physical instability to arise during the bouncing epoch.

Brane-world models permit a new approach to these difficulties,
since they appear to allow the possibility that brane-bound
observers might experience a bouncing universe, while embedded
within a stable higher-dimensional geometry 
\cite{prodanov,coule,gregory,marco,varum,myung,mukherji,yiota,ponce,china,myers}. 
Ref.~\cite{marco}
pointed out  such apparent example, consisting of a four
dimensional cosmology based on a brane world embedded in a
5-dimensional Reissner-Nordstr\"om 
background\footnote{See also the earlier works 
\cite{coule,gregory}.}. In this case, a
solution for the appropriate junction conditions can be provided
explicitly, with the result that the apparent
weak energy condition violation term arises from the projection of
bulk curvature effects onto the brane. This construction has
recently been criticized, however, as inheriting the instability
to fluctuations of the underlying Reissner-Nordstr\"om geometry
\cite{myers}.

Our purpose in this paper is to broaden the context of the
discussion, by presenting a number of new brane-world
constructions for which the brane-bound observers experience
bounces. In particular we do so for geometries which do not
require bulk electric fields, and which may not have the same
stability problems which afflict the Reissner-Nordstr\"om example.
We provide two classes of examples, which differ according to
whether or not the relevant brane is a probe brane moving in a
bulk spacetime, or is a boundary brane subject to an appropriate
set of Israel junction conditions.

We present our results in the following way. In the next section
we review the usual conditions which are required in order for
four-dimensional FRW cosmologies to bounce.
In particular, we focus on the case of universes with negative or
flat curvature, in which the bounce requires a violation of the
weak energy condition, leading to non-standard acceleration.
Section 3 then
describes several models with  bounce having boundary 3-branes moving in
various five-dimensional geometries. These examples include simple
$S$-brane-like geometries which are solutions to the Einstein
equations, as well as examples which also involve bulk dilatons
and gauge fields. The resulting background  geometries typically
have either time-like or space-like singularities as well as
Cauchy horizons.
For each model, we identify the terms that appear in the effective
four dimensional Friedmann equation, which are responsible for the
bouncing behavior.
Section 4 presents similar models based on
observers riding probe branes within bulk spaces whose dimension
can be higher than 5. These move through field configurations
which solve higher-dimensional supergravity equations, and which
represent the field sourced by a stack of source branes. Being
supersymmetric, these bulk configurations are stable. Finally, our
conclusions are summarized in section 5.

\section{Bouncing and Acceleration in Standard Cosmology}\label{acstcos}
We start by reviewing the conditions for a bouncing behavior which
arise within 4D standard cosmology (see also, for example, \cite{visser}).
We express these requirements
in terms of the various energy conditions which the energy density
in the universe must satisfy (or violate) in order to obtain a
bounce. As we shall see, the same energy conditions also turn out
to be quite useful for understanding bouncing behaviors in
non-standard cosmologies based on extra-dimensional models.

For our purposes, the universe bounces when it passes smoothly
from a contracting to an expanding phase without encountering an
intervening singularity. That is, the universal scale factor,
$a(t)$, reaches a positive, minimum value $a_0 > 0$ in the past,
at a time which we choose to be $t = 0$. This turning point
separates contraction from expansion, and a necessary condition
for its existence is that the universe experiences positive
acceleration, $\ddot a > 0$, for some interval around the turning
point. The conditions for bouncing are consequently intimately
related to those for positive acceleration.

\subsection{Basic equations and energy conditions}
In standard cosmology, a metric which preserves the observed
homogeneity and isotropy in the three spatial dimensions has an
FRW form, and we restrict ourselves to this form from now on:
\be\label{FRWmetr}
    ds^{2}=-d \tau^{2}+a^2(\tau)
    \, d \Omega^{2}_{k,3}\,.
\ee
Here $d\Omega^2_{k,3} = dr^2/(1 - k r^2) + r^2 \, d\Omega_2^2$ is
the usual metric on the spatial slices, with $d\Omega^2_2$
denoting the line element on the 2-sphere. The constant $k$ takes
one of the three values $k = 0$ or $\pm 1$ for flat, spherical or hyperbolic
spatial topologies.

Homogeneity and isotropy imply the stress-energy density of matter
must have the perfect-fluid form:
\be\label{perfect}
    T_{\mu}^{\,\,\nu}= \hbox{diag}\, \left(-\rho,\,p,\,p,\,p \right)\,,
\ee
with, as usual, $p$ and $\rho$ being the fluid's pressure and
energy density. We assume these to be related by an equation of
state of the form
\be\label{simplstate}
    p\,=\,\omega \rho \,
\ee
with the parameter $\omega$ being constant.

Assuming the metric evolution to be governed by Einstein's
equations, the cosmological evolution of the the universe is then
given by the usual three equations,
\begin{itemize}
\item The equation of conservation of energy:
\be\label{conservenergy}
    \dot{\rho}+3\,H\,(\rho+p)\,=\,0\,\,,
\ee
\item The Friedmann equation
\be\label{normalfriedman}
    H^{2}+\frac{k}{a^2}=\frac{8\pi G_{4}}{3} \rho
\ee
\item The Raychaudhuri equation
\be\label{raychau}
    \frac{\ddot a}{a} = -\frac{4\,\pi G_{4}}{3} (\rho+3 p) \,.
\ee
\end{itemize}
Here $G_4$ denotes the four-dimensional Newton's constant and the
Hubble parameter is, as usual, $H \equiv {\dot{a}}/ {a}$. These
equations clearly tie the cosmic acceleration of the universe to
the energy and pressure of the matter it contains. We now
summarize several energy conditions which have been proposed to
characterize reasonable general conditions for how the pressure
and energy of stable matter can behave \cite{hawking,carroll}.

\begin{itemize}
\item {\it The weak energy condition (WEC)}, states that
$T_{\mu \nu} t^{\mu} t^{\nu}\geq 0 $ for any time-like vector
$t^{\mu}$. This condition states that energy density measured by
any observer is non-negative. Equivalently, in terms of energy
density and pressure, $\rho \ge 0$ and $\rho+p \ge 0$. The
validity of WEC is  in general crucial for the proof of
singularity theorems.
\item {\it The dominant energy condition (DEC)}, requires the same
constraints as does the WEC, with in addition the requirement that
$T^{\mu \nu} t_{\mu}$ cannot be space-like. This translates into
the condition $\rho \ge |p|$ or $1 \ge \omega$. When the DEC is satisfied, the
conservation theorem by Hawking and Ellis applies \cite{hawking},
which shows that energy cannot propagate outside the light cone
(and so, in particular, energy-momentum cannot spontaneously
appear from nothing).

\item {\it The strong energy condition (SEC)}, requires that
$T_{\mu \nu} t^{\mu} t^{\nu} \ge \frac{1}{2} T_{\lambda}^{\lambda}
\, t^{\sigma} t_{\sigma}$ for all time-like vectors $t^{\mu}$.
Equivalently, $\rho+p \ge 0$ and $\rho+3 p \ge 0$.
\end{itemize}

\vskip 0.4 cm
\subsection{Conditions for Bouncing}
A bounce requires $\dot a$ to pass smoothly from negative to
positive values, and so it follows that $H$ must vanish at some
(turning) point, around which $\ddot a > 0$ and $a > 0$.
Eq.~\pref{raychau} shows that $\ddot a /a > 0$ implies $p <
-\rho/3$, and so a bounce necessarily implies a violation of the
SEC.

Similarly, given the Friedmann equation,
eq.~\pref{normalfriedman}, the vanishing of $H$ implies conditions
on the energy density during the bounce. The precise condition
depends on the curvature, $k$, of the three-dimensional spatial
slices, as we now summarize.

\ssubsubsection{Bouncing conditions when $k=1$:}
This option is a textbook case, and in this instance a bounce
requires the energy density to violate the strong energy condition
(but not necessarily the WEC since $\rho$ need not pass through
zero even if $H$ does). So in this case the condition $p <
-\rho/3$ suffices, and no additional conditions need be imposed on
$\rho$. (For the equation of state, \pref{simplstate}, the
condition $p < -\rho/3$ implies $\omega < -\, \frac13$.) The
classic example of this type is furnished by the de Sitter
universe, for which $-p = \rho > 0$.

\ssubsubsection{Bouncing conditions when $k=0,-1$:}
This case is more difficult to achieve and so is less well
studied. It is not enough to have $p < -\rho/3$ to have a bounce,
because the energy must also pass through zero (if $k = 0$) or
become negative (if $k = -1$). In addition to violating the SEC, a
bounce also requires the energy density must violate the WEC (and
consequently also the DEC). It is noteworthy that once $\rho < 0$
the condition $p < -\rho/3$ becomes {\it weaker}, since it allows
positive $p$. (For instance, for the equation of state,
\pref{simplstate}, $\rho < 0$ and $p < -\rho/3$ implies $\omega >
-\, \frac13$ rather than $\omega < - \, \frac13$.)

Since the kinematics of this kind of bounce is less familiar, we
next examine this case in more detail, assuming an equation of
state, (\ref{simplstate}), with constant $\omega$. Energy
conservation, (\ref{conservenergy}), for these theories gives rise
to the relation
\be\label{consenergy}
    \rho=\rho_{0} \, \left(\frac{a(\tau)}{a_{0}} \right)^{-3(1+\omega)}
    = - c\cdot\,a^{-3(1+\omega)}(\tau) \,,
\ee
where $c = -\rho_{0}/a_{0}^{-3(1+\omega)}$. Since we imagine
$\rho_0 < 0$ at the turning point, we take $c > 0$.

If we assume for concreteness that $k = -1$, the Friedmann
equation can be written as
\be\label{simpboun}
    \dot{a}^{2} = 1- c\, a^{-(1+3 \omega)}\,.
\ee
where $c$ is the parameter defined in (\ref{consenergy}). From
this equality it is clear that the scale factor never becomes
smaller than $a_{min} = c^{1/(1+3\omega)}$, and $\dot a = 0$ when
$a = a_{min}$. It is not difficult to see that the universe
bounces at this minimum value. For example, in the special case
$\omega = \frac13$, one can solve eq.~(\ref{simpboun})
analytically to find
\be
    a^2(\tau) = \tau^{2} + c\,.
\ee
The resulting space-time is not singular for any finite $\tau$.

The Raychaudhuri equation, for the system we are considering,
takes the form
\be\label{rayspec}
    \frac{\ddot a}{a} = \frac{4\,\pi \, c\,  G_{4}}{3} (1+3 \omega)
    \cdot\,a^{-3(1+\omega)}(\tau) \,,
\ee
from which it is clear that $\omega > -\, \frac13$ ensures the
acceleration is always positive (although it asymptotically
vanishes).

Not surprisingly, the causal structure of a bouncing universe such
as this $k = -1$ example differs from the usual situation for
expanding FRW cosmologies.
In particular, an observer in the bouncing cosmology does not have
a
horizon which limits the distances from which she can
receive signals. By definition, an observer has a
horizon when the following integral converges:
\be\label{delta}
    \Delta \equiv \int_{t_{0}}^{+\infty} \frac{d t}{a(t)} < \infty\,,
\ee
since $\Delta$ represents the maximal distance that an observer at
time $t_{0}$ can probe. In the above $k=-1$ example, using
(\ref{normalfriedman}), we can write
\be\label{hori}
    \Delta = \int_{t_{0}}^{+\infty} \frac{d t}{a(t)} \\
    = \int_{a_{0}}^{+\infty} da \, \frac{a^{{(-1+3 \omega)}/{2}}}{\sqrt{
    a^{1+3 \omega}+c}} \,.
\ee
It is simple to check that this expression diverges, regardless of
the sign of $c$, provide $\omega>-\frac{1}{3}$. Consequently we
obtain a system having eternal positive acceleration (although it
vanishes asymptotically) without  horizons.

This behavior is in marked contrast with the eternal
positive acceleration that one meets in a de Sitter universe.
 The presence of horizons in the de Sitter case
has been argued to mean that the total number of degrees of
freedom contained in a causal patch of the universe is always
finite, an observation which is regarded as having problematic
consequences for field theory and string theory
\cite{fischler,susskind}. We see that, for $k=-1$, we have
acceleration without this horizon problem, and we provide in the
next section, brane world examples with this characteristic.

\subsection{Bouncing Brane Cosmologies}
The new ingredient which branes introduce into the bouncing
discussion is the potential they bring for separating the
requirement for negative energies from the necessity for having an
instability. They do so basically because there is more than one
metric which appears in the physics of observers on a brane, and
the metric which shows the bounce need not be the metric which
appears in the Einstein action. We briefly summarize these issues
here, before discussing several examples of bouncing branes in
subsequent sections.

The distinction between the various metrics which are relevant to
brane observers is most starkly highlighted in the situation where
a brane moves through a static bulk spacetime --- the so-called
`mirage cosmology' \cite{kiritsis,mirage}. In this case the induced metric,
$\gamma_{\mu\nu}$, on the brane is typically time-dependent,
because of its dependence on the time-dependent brane position,
$y^M(t)$:
\be
    \gamma_{\mu\nu} = g_{MN} \, \partial_\mu y^M \, \partial_\nu y^N
    \,.
\ee
Since it is the induced metric which governs the propagation of
particles living on the brane (like photons), for all observations
performed by brane observers using these particles this time
dependence would be indistinguishable from what would be seen from
a stationary brane sitting in a time-dependent bulk spacetime.

On the other hand, the metric which appears in the low-energy
effective 4D Einstein equation is the lowest Kaluza-Klein mode of
the bulk metric, $g_{\mu\nu}$, and this in general differs from
$\gamma_{\mu\nu}$. A brane observer can detect the difference
between these two metrics since this difference generates a host
of preferred-frame effects, which can be constrained even within
the gravitational sector \cite{LV}.

In this situation a new interpretation of the Friedmann equation
becomes possible, and this is what allows the presence of
negative-energy terms in it to be potentially benign. To this end
consider the interpretation of a brane observer who is unaware of
the difference between the two metrics, and who (like us) simply
sees particles like photons propagating within a time-dependent
metric. For isotropic and homogeneous spacetimes, the resulting
blue/redshifts of the photons would be interpreted as being due to
an FRW cosmology whose Hubble parameter, $H$, could be computed.
Not knowing the distinction between $\gamma_{\mu\nu}$ and
$g_{\mu\nu}$, this observer would interpret any dependence of
$H^2$ on the scale factor, $a$, as being due to the presence of
various forms of energy density, through the Friedmann-like
formula \pref{hubblebran}. In particular, if the induced metric
for this observer bounces, the observer would believe there must
be negative energy present, as discussed above. The brane observer
could similarly deduce the pressure of the various cosmological
fluids using either conservation of energy or the Raychaudhuri
equation \cite{kiritsis}.

The bulk observer would recognize that there is no real
negative-energy field associated with the apparent bounce as seen
by the brane observer, and so there also need not be any
instability associated with it. Recognizing that there are two
metrics in the problem, the bulk observer would see that these
permit the definition of two scale factors, $a_g$ and $a_\gamma$.
Although the bulk Einstein-Hilbert action implies the Friedmann
equation in the usual way for $a_g$, it is only an `effective'
Friedmann equation for $a_\gamma$ which the brane observer
determines from photon observations.\footnote{The brane observer
could draw similar conclusion given sufficiently accurate
measurements, particularly using gravitational waves.} (We call
the Friedmann equation for $a_\gamma$ `effective' because the
kinetic term for $\gamma_{\mu\nu}$ comes partially from the brane
kinetic energy, rather than purely from the Einstein-Hilbert
action.)
What the brane observer would say is a strange type of dark energy
which does not couple to the visible fields on the brane, the bulk
observer understands to be an artifact of the brane observer's
making observations using particles which are confined to a moving
brane universe;
 on the other hand, the bulk observer is able to interpret these brane terms
as due  to particular properties of the higher dimensional system,
that is not possible to realize at the level of projected lower dimensional
physics.

It is this loop-hole which we wish to explore with the examples we
provide in this paper. We also believe that the analogy between
the misguided brane observer and the present-day evidence for dark
energy is sufficiently uncomfortable to warrant a more systematic
study of preferred-frame and gravity-wave effects in post-BBN
 cosmology.

\section{Bouncing Boundary Branes}
We have seen that for $k = 0,-1$ cosmologies --- which are often
the ones of current interest --- bouncing requires a violation of
WEC and of DEC (which is to say vanishing or negative energy
density for some observers).
In the recent literature, systems that violate the DEC have been
considered (see for example \cite{caldwell}), but they often
result,
 in instabilities \cite{carroll}.

The new feature of brane-world models is the possibility which
they raise of allowing bounces without necessarily paying the price of
instability. Although observers on the brane do see a cosmological
`fluid' which violates DEC, this fluid is the projection of bulk
curvature onto the brane. And so long as the full bulk theory is
stable, it may be that this 4D DEC violation need not necessarily
imply an instability of the full theory. Our goal in the remainder
of this paper is to provide examples of brane-world models using
spacetimes which appear to be stable.

The original proposal along these lines argues that an observer
confined on a brane embedded into a five dimensional
Reissner-Nordstr\"om geometry experiences a $k = 0,\pm 1$ bounce for
some brane trajectories \cite{marco,yiota}. This observer finds
source terms in her effective 4D Friedmann equation which appears to
have negative energy, in agreement with the arguments of the
previous section. Seen from the 5-dimensional perspective, these
terms arise as bulk curvature contributions which are projected
onto the brane. For the Reissner-Nordstr\"om example they are
nonzero when the 5D black hole carries nonzero charge, and are
zero otherwise. For vanishing charge the WEC-violating terms
disappear, and the model has an initial singularity (at least when
$k = 0$ or $-1$).

The purpose of presenting these examples is to broaden the
discussion of the connections between bounces, singularities and
instabilities. For instance, bulk stability has been partially
studied for some of the $S$-brane-like examples we provide in this
section, with some indications that they may not share the bulk
instabilities of the Reissner-Nordstr\"om geometry. The examples
we provide also show that boundary-brane bounces are possible for
pure gravity (with a cosmological constant) and so the presence of
an electric charge in the background is not required. Instead,
 a feature which all of the examples of this section share
with the Reissner-Nordstr\"om example is the presence of  naked (time
or space-like) 
singularities in the bulk spacetime, and this makes it difficult
to rule out instabilities which arise due to fluctuations which
propagate out of these singularities and reach the brane.
 This is likely related to the violation of the dominant energy
condition (DEC) which the 4D observer sees, since this energy
condition is used in the proof that energy and momentum cannot
appear acausally, from outside the observer's light cone.

In the remainder of this section we first review the origin of
these WEC-violating contributions to the effective Friedmann
equation, and then use the results to examine several new
co-dimension one brane-world scenarios.

\subsection{Boundary Branes and the Effective Friedmann Equation}\label{formabw}
Consider a five-dimensional bulk space time, $M$, that contains
($3+1)$-dimensional branes, $B$, on which 4-dimensional observers
may be forced to live. We further imagine that these branes are
boundary branes, $B = \partial M$, which we realize by imagining
taking two copies of $M$ on one side of $B$, and gluing these
together at $B$ to define a covering space for which the two sides
of the brane are identified by a ${\mathbb Z}_{2}$ symmetry.

The back-reaction of the brane onto the geometry of the spacetime
is obtained by requiring the system to satisfy appropriate
matching conditions which relate the stress energy on the brane to
the discontinuity of the geometry across the brane. The relevant
discontinuity is in the extrinsic curvature, ${\mathcal K}_{M N} =
\nabla_{M} \eta_{N}$, where $\eta_N$ is the unit normal to the
brane. The discontinuity condition states \cite{israel}
\be\label{israel}
    \Delta {\mathcal K}_{\mu \nu} \equiv {\mathcal
    K}_{\mu  \nu}^{+} -{\mathcal K}_{\mu \nu}^{-}
    =-\kappa_{5}^{2}\,\left(\tilde{T}_{\mu \nu}-\frac{1}{3}
     \tilde{T}^{\lambda}_{{}\lambda} \gamma_{\mu \nu}
    \right)\,,
\ee
where $\tilde{T}_{\mu \nu}$ denotes the stress energy of matter on
the brane, and $\gamma_{\mu \nu}$ is the induced metric
on the brane.

Explicitly, consider the following ansatz for the five-dimensional
metric
\be\label{ansatz}
    ds_{5}^{2} = -h(r)\, d t^{2} +
    \frac{d r^{2}}{g(r)}+r^{2}d\, \Omega_{k,3}^{2} \,,
\ee
where as before $d\, \Omega_{k,3}^{2}$ represents the measure of a
maximally symmetric three-dimensional subspace of constant
curvature $k$. Different kinds of horizons for this metric correspond to
null surfaces, $r = r_{h_i}$, along which $g(r)$
vanishes (provided these are not also curvature singularities).
When such horizons exist $h$ and $g$ typically change sign, and
this can change the geometry's Killing vectors from being spacelike to
being timelike (or vice-versa). In particular the metric is static (with
$t$ as the time coordinate) if $h$ and $g$ are both positive, but
it is explicitly time-dependent (with $r$ as the time coordinate)
if $h$ and $g$ are both negative.

A simple choice for a 3-brane position is along a surface $r =
r_b(t)$. Given this choice the induced four-dimensional metric is
then completely specified to be
\bea
    d s^{2}_{induced} &=& - \left[ h(r_{b})
    - \frac{1}{g(r_b)} \left( \frac{dr_b}{dt} \right)^2 \right] \,
    dt^2 + r_b^2 \Omega_{k,3}^2 \nonumber \\
    &=&
    -d \tau^{2}+a(\tau)^{2}d\, \Omega_{k,3}^{2} \,,
\eea
from which we see the brane's proper time is given by $d\tau /dt =
\left[ h - ({r'}_b^2/g)  \right]^{1/2}$, the induced scale factor
is $a(\tau) = r_b[t(\tau)]$ and so the resulting Hubble parameter
is $H = \dot a/a = (r'_b /r_b) \left[ h - ({r'}_b^2/g)
\right]^{-1/2}$. Notice that the derivative ${r'}_b = {dr_b}/{dt}$
vanishes if the brane sits at a fixed coordinate position, $r =
r_b$, in a static geometry and so $H$ vanishes in this limit. On
the other hand, $r'_b \to \infty$ if the brane sits at a fixed
coordinate (spacelike) position, $t = t_b$, in a time-dependent geometry, and
in this case $H \to |g|^{1/2}/r_b$.

Since ${r'}_b$ determines the brane's extrinsic curvature, the
trajectory, $r_{b}(t)$, is fixed in terms of the stress energy on the
brane by the junction conditions, \pref{israel}. Thus, the
cosmological evolution in four dimensions, obtained by solving the
junction conditions at the singular surface, has a
clear geometrical interpretation in the motion of the brane along
a time-like trajectory in the higher dimensional background
\cite{mirage}.

\smallskip
Let us use the junction condition (\ref{israel}) to determine the
form of the effective Friedmann equation as seen by an observer
riding on the brane in our background. The extrinsic curvature can
be written as
\be\label{extrinsictwo}
    {\mathcal K}_{\mu \nu}=\frac{1}{2} \eta^{\sigma} \
    \frac{\partial \gamma_{\mu\nu}}{\partial x^{\sigma}}
\ee
and, in our case, we take $\eta^{\sigma}=\pm\,(\,(h
g)^{-1/2}\,\dot{r}_{b}, \,(h g)^{1/2}\,\dot{t},\,0,\,0,\,0)$,
and $\dot{r}$ and $\dot{t}$ are derivatives
with respect the proper time $\tau$. There are two nontrivial junction
equations arising from the time and space components of (\ref{extrinsictwo}).
The  spatial components of the extrinsic curvature can be found to be
\be\label{extrinsicthree}
    {\mathcal K}_{i j}\,=\,r_{b}\,g^{1/2}
    \,\sqrt{1+\frac{\dot{r}_{b}^{2}}{g}} \,
    \gamma_{ij}\,,
\ee
When the energy density on the brane has a perfect fluid form, and there
is no direct coupling between bulk fields and brane matter besides gravity,
the conservation of energy momentum tensor on the brane lead to the
the energy conservation  equation (\ref{conservenergy}) for the energy
density on the  brane~\footnote{The situation
 in which there is a direct coupling
between bulk fields and brane matter will be discussed in specific examples
in the next sections.}.
 It turns out that the time component of the extrinsic curvature,
 gives a third, non independent equation, so we need only to consider
the spatial components as well as the energy conservation equations as
evolution equations.

Assuming, again, that  there is no direct coupling between bulk fields
(besides gravity) and brane matter, the spatial components of the
Israel junction conditions, obtained from (\ref{extrinsicthree}),
gives us the effective Friedmann equation on the brane:
\be\label{frie}
    \frac{\dot{a}^{2}}{a^2} =\frac{\kappa_{5}^{4}\, \rho^{2}}{36}
    -\frac{g(a)}{a^{2}}\,,
\ee
As it is clear from this formula, it is the
term proportional to $g(a)$ which can give rise to negative
contributions to the energy as seen by a brane observer.

It is useful to write eq.~\pref{frie} in a way which is more
suitable for both a numerical or qualitative approach. Writing
\be\label{hamilton}
    \dot{a}^{2}+W(a)=0\,, \ee where \be\label{potential} W(a) \,=\,
    g(a)- \rho^{2}\,a^{2}\,,
\ee
the Friedmann equation is reduced to a Hamiltonian constraint for
a particle having zero energy~\cite{myers}. The potential $W(a)$
must be non-positive in order to have a solution of
(\ref{hamilton}). It turns out that the solution has a bounce if
$W$ vanishes for some scale factor, $a_t$, greater than zero,
corresponding to a classical turning point of the particle motion.
To decide if a cosmological model bounces, it suffices to plot the
function $W$. This also provide us
with constraints on the model parameters which have to be satisfied in order
to get a bounce.

In the next sections, we present various new examples of
brane-world models with a  bouncing behavior. Our aim is mainly to
describe the bouncing phase of the history of these universes; for
this reason, in general we do not discuss the subsequent
cosmological evolution in these models.

\subsection{Pure Gravity S0-Brane Examples}\label{efirex}
In this section we provide additional examples of bouncing,
co-dimension 1 brane worlds which are similar in spirit to the
Reissner-Nordstr\"om proposal of ref.~\cite{marco,yiota}. A motivation
for examining more solutions comes from the recent observation
that the Reissner-Nordstr\"om brane-world model shares the
instability of the underlying 5D geometry to back-reaction from
fluctuations near the black-hole horizon \cite{myers}. As such it
cannot decide the stability issue.

Consider first five-dimensional gravity with a negative
cosmological constant,
\be
    S_5=\int d^5 x \, \sqrt{-g}\left[\frac{1}{2 \kappa_{5}^{2}}
     R + 12 \,\Lambda \right] \,  + S_{4}\,,
\ee
where $S_{4}$ is the action of a four brane embedded in the full
five-dimensional space-time and $\Lambda$ is positive. The
solution of interest to Einstein's equations in the bulk is
\be\label{simsbrmetric}
    ds_{5}^{2}=-h(r)\, d t^{2} +
    \frac{d r^{2}}{h(r)}+r^{2}d\, \Omega_{-1,3}^{2}
\ee
with
\be\label{fsolSb}
    h(r)= -1+\frac{\xi^{2}}{r^{2}} +\Lambda r^{2}\,,
\ee
where $\xi$ is an integration constant which is physically
interpreted as measuring the tension of the time-like
singularities which source the gravitational field. Notice that
the surfaces of constant $t$ and $r$ are negatively curved.

If $4\Lambda \xi^2 < 1$ this geometry has two horizons, at $r =
r_\pm$ with $2\Lambda r_\pm^2 = 1 \pm \sqrt{1 - 4\Lambda \xi^2}$.
Writing $h(r) = \Lambda(r^2 - r_+^2)(r^2 - r_-^2)/r^2$ shows that
$h > 0$ (and so the metric is static) for $r > r_+ > r_-$ and $r <
r_- < r_+$, but $h < 0$ (and so the metric is time-dependent) if
$r_- < r < r_+$. The resulting causal structure can be represented
by a Penrose diagram which looks like that of an
AdS-Reissner-Nordstr\"om black hole. For $\Lambda =0$, this
background corresponds to an S0-brane geometry largely discussed in
\cite{bqrtz,qtz,bmqtz}.

The effective Friedmann equation for an observer bound to a brane
moving along a trajectory $r = r_b(t)$ \footnote{The first example
of brane-world in this geometry for zero bulk cosmological
constant was presented in \cite{gqtz}.} (or $t = t_b(r)$) in this
geometry takes the form
\be\label{nsfe}
    H^{2} = \rho^2 - \frac{h(a)}{a^{2}} \,,
\ee
where $a(\tau)= r_b(\tau)$ and  $\kappa_{5}^{2}=6$.
Writing the energy density on the brane as $\rho = \lambda +
\rho_{m}$ with $\lambda$ constant, eq.~(\ref{nsfe}) becomes
\be\label{frinef}
    H^{2}=2\lambda \rho_{m}+ \rho_{m}^{2}+(\lambda^{2}-\Lambda)
    +\left[ \frac{1}{a^{2}}-\frac{\xi^{2}}{a^{4}} \right]
    \,.
\ee
We identify the terms appearing in the RHS of (\ref{frinef}) as
follows. The first term has to form of the standard, linear term
in $\rho_{m}$, provided we define an effective Newton's constant
in terms of the brane tension. The second is the now-familiar
quadratic correction in $\rho_m$, the third term, between normal
parenthesis, comes from the cosmological constants in the bulk and
on the brane. The fourth term is the usual contribution that comes
because our spatial slices have negative curvature. The final term
is induced from the bulk and is the one which is most important
for our discussion, since it looks like a negative contribution to
the energy which scales with $a$ like radiation ({\it i.e.} with
$\omega = \frac13$). This term is precisely of the form discussed
in Section (\ref{acstcos}) which is required if the brane observer
is to see a bounce.

We now concentrate on two interesting special cases. The first is
the simplest brane-world model that gives a bounce, consisting of
a tensionless brane embedded in an empty bulk. The second is a
more realistic model that contains energy density on the brane in
the form of radiation.
In both these two examples,
an observer on the brane probes physics in a frame in which the Planck
mass is constant in time.

\subsubsection*{A Simple Special Case}
As our first example,  consider a tensionless, empty brane
embedded in a pure S-brane background. Since we seek simple
examples of bounces rather than realistic descriptions of the
present-day universe, we take $\Lambda = \lambda = \rho_m = 0$ in
the previous example. The absence of stress-energy on the brane
ensures the brane world-sheet must embed into spacetime with
vanishing extrinsic curvature, and inspection of
eq.~\pref{extrinsicthree} shows that this is only possible if
$h(r_b) < 0$ and $\dot r^2 = |h(r_b)|$. This corresponds to the
brane lying in the time-dependent regions of the bulk, sweeping
out a surface of constant spatial coordinate, $t_b$. The Penrose
diagram for such a brane-world is shown in Figure \ref{caso1}.

In this case, eq.~(\ref{nsfe}) reduces to
\be\label{nsfe1}
    H^{2} =+\frac{1}{a^{2}}- \frac{\xi^2}{a^{4}}\,,
\ee
which can be integrated analytically, leading to
\be\label{ssl}
    a^{2}(\tau)=\tau^{2}+\xi^{2}\,.
\ee
The acceleration of the scale factor is everywhere positive,
vanishing only asymptotically:
\be\label{accSbr}
    \frac{\ddot{a}}{a}=\frac{\xi^{2}}{(\tau^{2}+\xi^{2})^{2}} \,.
\ee
The resulting cosmology has positive acceleration at any finite
time, without the  horizon problem  as discussed after formula
(\ref{hori}).

{}From these expressions it is clear that the scale factor never
vanishes, so the 3-brane has a smooth bounce that crosses the
intersection point between the Cauchy horizons. This bouncing
geometry may be traced to the time-like singularities which source
the bulk metric, since it is the non-standard term proportional to
$\xi^{2}$ which is responsible. If the singularities are
eliminated by sending $\xi \to 0$, the model becomes a standard
negative-curvature-dominated empty universe, with the usual
singularity at the origin.

\FIGURE[h]{
\let\picnaturalsize=N
\def\picsize{2.8in}
\def\picfilename{sbranebounce.eps}
\ifx\nopictures Y\else{\ifx\epsfloaded Y\else\input epsf \fi
\let\epsfloaded=Y
\centerline{\ifx\picnaturalsize N\epsfxsize \picsize\fi
\epsfbox{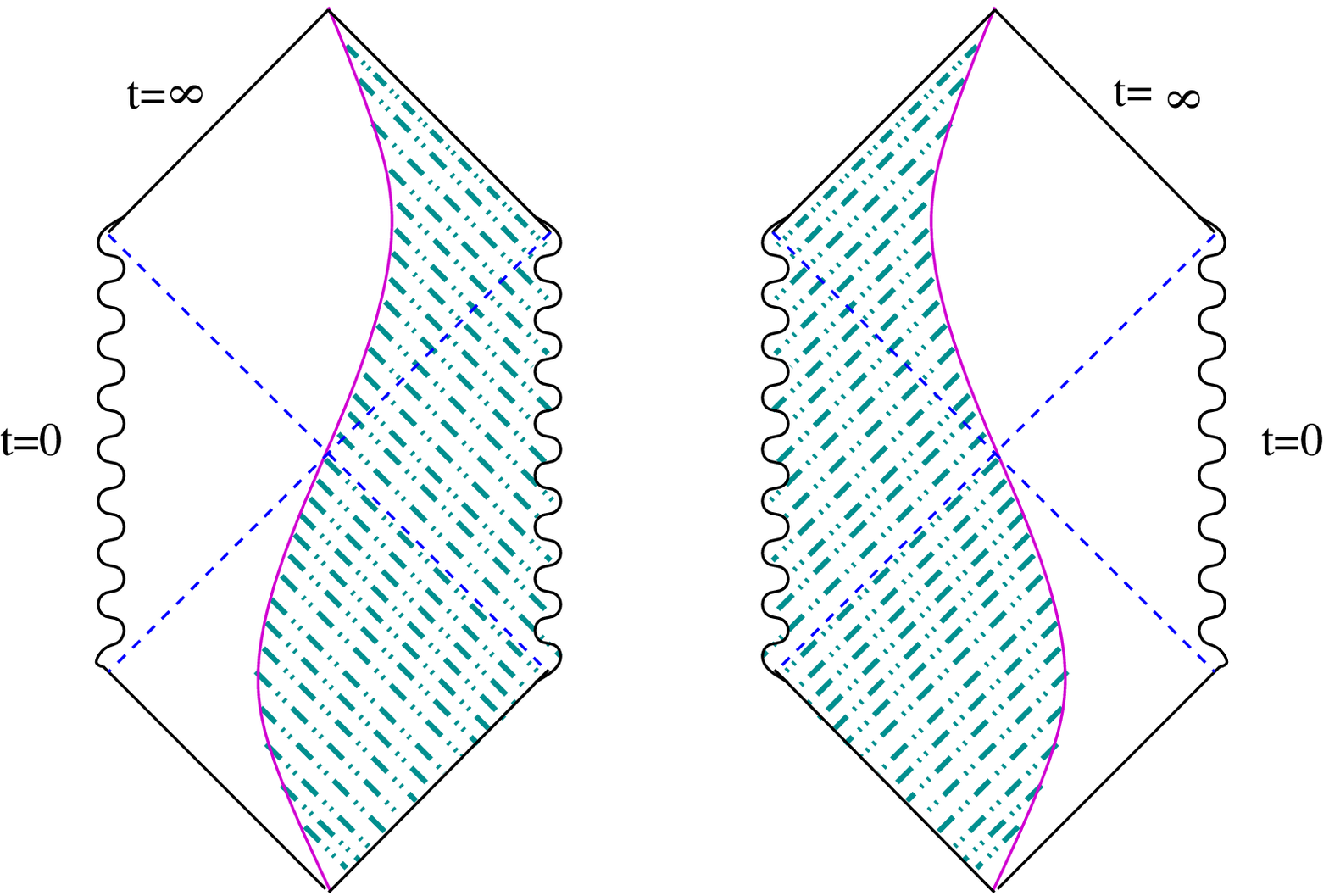}}}\fi
\caption{Graphical representation of a brane-world embedded in an S-brane
background. \label{caso1}}}

\medskip
\noindent {\it Stability against Perturbations}
\medskip

\noindent
The stability of the bulk spacetime has been studied with mixed
results. Although ref.~\cite{bqrtz,bmqtz} found the geometry to be
unstable to general perturbations, ref.~\cite{costa} has pointed
out that it is stable if fields are assigned initial conditions
having compact support in the remote past. It is not yet known
whether physical considerations can decide which set of boundary
conditions should be used, or if instabilities may be generated by
perturbations which propagate in along the horizon from the
time-like singularities.\footnote{We thank Rob Myers for
conversations on this point.}

The same issues arise when the boundary brane is present, and
these ensure the spacetime is stable against fluctuations which
are confined to the brane (since these have compact support in the
past, when viewed from the bulk). To see this explicitly let us
evolve the fluctuations of a brane-bound massless,
minimally-coupled scalar field, $\phi$, using the induced brane
metric, whose amplitude is small enough that it does not
significantly change the cosmological evolution of the background
\cite{marco}.

Using conformal time, the induced metric seen by the scalar field
is
\bea
    d s_{4}^{2} &\equiv& a^2(\eta) \left[-d \eta^{2}
    + d \Omega_{-1,3}^{2} \right] \nonumber \\
    &=& \xi^2 \cosh^{2}{\eta} \left[-d \eta^{2}
    + d \Omega_{-1,3}^{2} \right] \,,
\eea
and the equation of motion for a fluctuation mode, $v_k = a\,
\delta \phi_{k}$, with a given co-moving momentum, $k$, becomes
\be
    v_{k}''+\left(k^{2}-\frac{a''}{a} \right)v_{k}=0 \,.
\ee
Here primes denote differentiation with respect to $\eta$, and for
the metric under consideration $a(\eta) = \xi \, \cosh \eta$, so
$a''/a = 1$.

The WKB solutions for modes not near $k^{2} = 1$, is
\bea
    \delta \phi_{k}^\pm = \frac{v^\pm_{k}}{a} &=& \frac{e^{\pm i \eta\sqrt{k^{2}-1}
    }}{ a(\eta) \left[2\sqrt{k^{2}-1} \right]^{1/2}} \qquad \hbox{if} \quad |k| > 1\,,
     \nonumber \\
    &=& \frac{e^{\pm \eta \sqrt{1 - k^{2}} }}{a(\eta) \left[2\sqrt{1- k^{2}}\right]^{1/2}}
    \, \qquad \hbox{if} \quad |k| < 1  \,.
\eea
Because $a(\eta) \propto  e^\eta$ for large $\eta$, these
expressions show that for all $k$, $\delta \phi^\pm_k(\eta)$
remains bounded for any $\eta$ (despite the exponential growth of
$v_k^+$ for $|k|<1$, which arises because of the positive
acceleration: $a''/a>0$). This is in particular true through the
problematic bouncing region, $\eta \simeq 0$, during which the
brane crosses the bulk-space horizon. Notice also that for large
$k$ the resulting power spectrum is $k^3 |\delta \phi_k|^2 \propto
k^{n-1}$ with $n = 3$, as is expected for a bouncing cosmology.

Although this calculation indicates an absence of instability
caused by brane-bound fluctuations, it does not address the
stability of the underlying spacetime to fluctuations along the
Cauchy horizon, and so cannot ultimately arbitrate the issue of
stability.

\medskip\noindent {\it Compactification}
\medskip

\noindent In scalar-tensor theories, it is well known that field
redefinitions can change a metric which bounces into another in
which the bouncing behavior is different, or even disappears
\cite{kachru}.
 Since it is a bounce in the Einstein frame (for
which the Einstein-Hilbert action has the standard
scalar-independent form) for which the analysis of Section 2
relies, it is important to check that the bounces we obtain are
really bounces when viewed from the 4D Einstein frame. Since the
example of present interest is so simple, it allows this to be
checked explicitly.

In the  example we are considering,
 the Planck mass is independent on time.
Let us ask what happens,
in this case, when  the spatial
extra dimension is compactified on a circle. Recall that (for
$\Lambda=0$) tensionless branes sit at a fixed spatial coordinate,
$t = t_b$, in the time-dependent region for which $r > \xi$ is the
time coordinate and $h(r) = -1 + \xi^2/r^2 < 0$. In this case,
since the geometry is invariant under translations of $t$, we may
compactify in this direction by identifying points for which $t
\sim t+L$, for some $L$. Once this is done, the dimensional
reduction from 5 to 4 dimensions introduces the time-dependent
circumference $C \propto \sqrt{|h(r)|} \, L$ into the
Einstein-Hilbert action, implying the compactification introduces
a time-dependence on the effective 4D Newton's constant.
Consequently, let us transform the metric to go to the Einstein
frame, in which the Planck mass is constant: we must rescale the
4D metric according to $g_{\mu\nu} \to g_{\mu\nu}/\sqrt{|h|}$, and
so repeating the above arguments leads to the scale factor:
\be
    a_E(r) = \frac{r}{|h(r)|^{1/4}} =
    \frac{r}{\left| 1 - \xi^2/r^2 \right|^{1/4}}\,.
\ee
Similarly, the  proper time becomes
\be
    \left( \frac{d\tau_E}{dr} \right)^2 = \left( \frac{1}{|h|}
    \right)^{3/2}
    = \frac{1}{\left| 1 - \xi^2/r^2 \right|^{3/2}}\,,
\ee
and so $\tau_E$ increases monotonically with $r$ as $r$ either
runs from $\xi$ to $\infty$ or from $\infty$ to $\xi$. The Hubble
parameter is therefore
\be
    H^2 = \left( \frac{da_E}{dr} \, \frac{dr}{d\tau_E} \right)^2
    = \frac{ \left[ 1 - 3 \xi^2/(2 r^2) \right]^2}{1 - \xi^2/r^2}
    \, .
\ee
We see that, in the Einstein frame, a bounce again occurs, but the
turning point is located at $r^2 = \frac32 \, \xi^2$, and so is
away from the horizon of the bulk geometry (which is situated at
$r = \xi$). For instance, if $r$ increases from $\xi$ to $\infty$,
$a_E$ initially contracts from an infinite value when $r = \xi$
until it reaches a minimum at $r^2 = \frac32 \, \xi^2$ and then
grows asymptotically as $r \to \infty$.

Notice that the above explicit compactification is possible for
the tensionless brane because it necessarily moves only within the
time-dependent regions of the bulk geometry. If the brane were to
pass into the static regions, spatial compactification would
necessarily have to be done in the $r$ direction along which the
metric is no longer symmetric.

\subsubsection*{An example with brane matter}
We next consider equation (\ref{frinef}) in more detail. Although
the general case is not amenable to analytic solution, it may be
treated using the method of the effective potential, as explained
in Section (\ref{formabw}). The effective potential relevant to
the general case \pref{potexII} is
\be\label{potexII}
    W(a) = -2\lambda \rho_{m}\,\, a^2 - \rho_{m}^{2}\, \,a^2
    - 1+\frac{\xi^{2}}{a^{2}}
    -(\lambda^{2}-\Lambda)\,\,a^2\,,
\ee
and our interest is in whether this function has zeros, which
represent the locations of a bounce.
 Let us consider a matter energy density, $\rho_m$, consisting of
radiation: $\omega = \frac13$. The potential \ref{potexII} becomes
\be
 W(a) = -\frac{c^2}{a^6} +\frac{(\xi^2-2\lambda\,c)}{a^2} -1 -
     \lambda_{4eff}\,a^2\,,
\ee
where we define $\lambda_{4eff}= \lambda^{2}-\Lambda>0$, and we
take this quantity to be positive. From the above expression it is
possible to argue that in order to have a bounce on the brane, we
must choose $\xi^2-2\lambda\,c > 0$.

\FIGURE{
\let\picnaturalsize=N
\def\picsize{2.5in}
\def\picfilename{cham2.eps}
\ifx\nopictures Y\else{\ifx\epsfloaded Y\else\input epsf \fi
\let\epsfloaded=Y
\centerline{\ifx\picnaturalsize N\epsfxsize \picsize\fi
\epsfbox{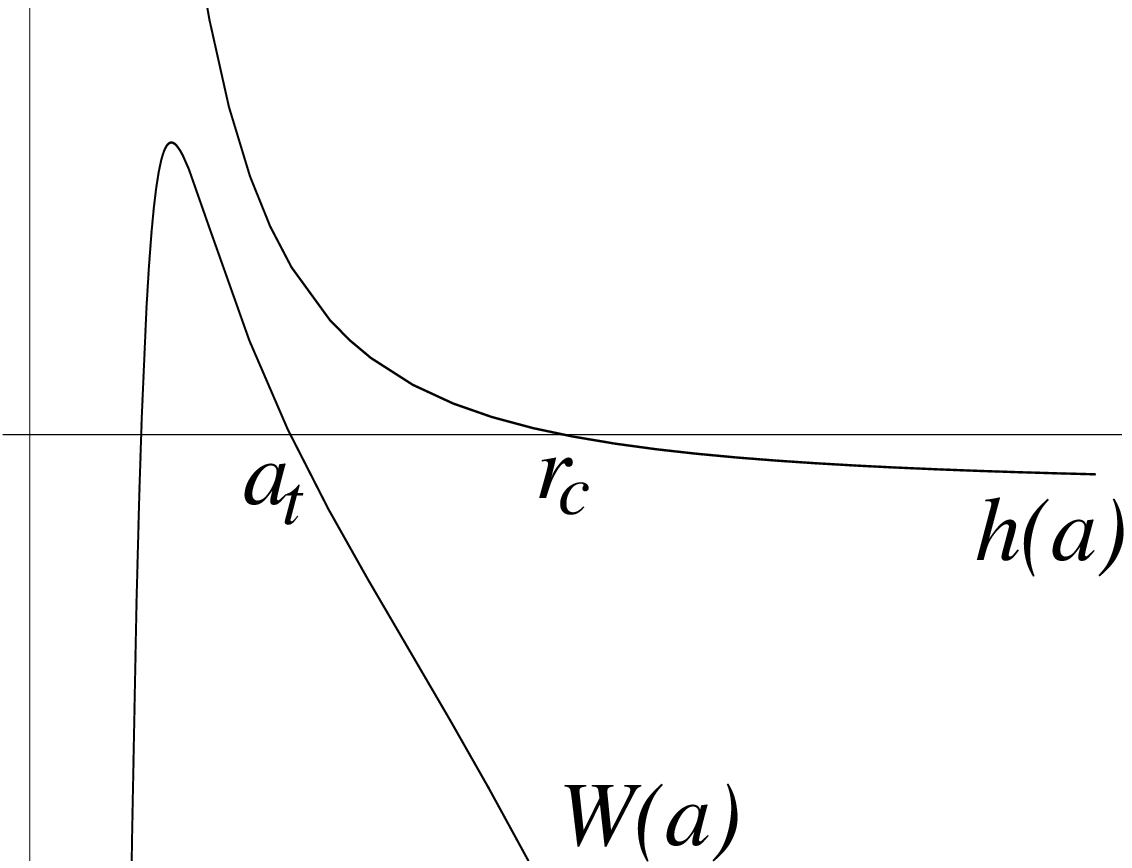}}}\fi
\caption{Plot of the potential $W$ of (\ref{potexII}) and of the
metric function $h$ in (\ref{fsolSb}). Writing
$\rho_{m}=c\,a^{-3(1+\omega)}$, we choose $\omega=1 / 3 $, $\xi
=3$, $\lambda=1$, $\Lambda=10^{-4}$, $c=1$.
 Notice that we have a bonce around the turning point
represented by $a_t$, the minimum value for the scale factor, that
corresponds to a zero for $W$. This point is inside the Cauchy
horizon, which is labeled by $r_C$.}\label{bo}}

Once this
 condition  is satisfied, there can be a bouncing  turning points for
certain values of the parameters. We plot in Fig.~\pref{bo} the potential,
$W(a)$,
as well as the metric function $h(a)$ that characterize our background for
some of these values, as indicated in the figure caption.
 As the
figure makes clear, $W$ has two zeros. The left-most zero
corresponds to a universe of finite lifetime with an initial and
final singularity, but the zero labeled $a_t$ corresponds to the
turning point of a bouncing  universe without
singularities. The corresponding brane-world
necessarily crosses the Cauchy horizon of the bulk geometry, which
is marked in the figure as $r_{C}$.

\subsection{S0-Branes with Dilatonic Bulk Matter}
In this section we consider more complex examples, where the bulk
solution contains a dilaton which we endow with a non trivial
Liouville-type potential, given by the sum of several exponential
terms. Again, our interest is in exhibiting simple bouncing
cosmologies rather than describing the present-day 
universe.


For this example we  consider the following five-dimensional
action, containing gravity and a scalar field having a bulk
potential given by a sum of two Liouville terms, plus a potential
on the 3-brane~\footnote{An example of  bouncing 
brane-world with a simpler dilaton potential can be found in 
\cite{foffa}.}. The purpose of this example is to understand how
the introduction of bulk matter changes the relationship between
the bounce and the horizons and singularities of the bulk
geometry,
 and how the choice of the bulk field potential is connected with
the bounce properties from a lower-dimensional point of view.
 We find that a direct coupling between the bulk scalar and the
energy density on the brane is necessary for the consistency of
the model. This fact makes the brane cosmology during the bouncing
phase non-standard, since there is a non-vanishing flow of energy
between bulk and brane.

The action for the model is\footnote{In this and next sections we
take $2\kappa_5^2=1$.}
\be
    S=\int d^{5} x \sqrt{-g} \left[ R-\frac{1}{2}
    (\partial \phi)^{2}
    -\Lambda_{1} e^{-{2}\phi/({3\sigma})}
    +\Lambda_{2} e^{-\sigma\phi}\right] +\int d^{4} x \sqrt{-g_{4}}\,
    \rho\, U(\phi)\,,
\ee
where $\rho$ denotes the energy density of all of the brane-bound
fields (including the tension and possible contributions from
various forms of matter). For now we leave the coupling function
$U(\phi)$ unspecified, although we shall later specialize to an
exponential function.

A solution to the coupled dilaton-Einstein field equations for
this bulk system was found in ref.~\cite{bnqtz}, and is given by
\bea
    ds_{5}^{2} &=& -h(r) \, d t^{2}+\frac{r^{3\sigma^{2}}}{h(r)}
    \, d r^{2} +r^{2} \, d \Omega_{k,3}^{2} \,, \nn \\
    \phi &=& 3 \sigma \ln r \,,
\eea
with
\be \label{ADSbrane}
    h(r)= -\frac{8k \, r^{3 \sigma^{2}}}{(3\sigma^{2}-2)(4+3\sigma^{2})}
    +\frac{2 \Lambda_{2} \, r^2 }{3(8-3\sigma^{2})} \,,
\ee
and where the constants $\sigma$, $\Lambda_i$ and $k = 0,\pm1$ are
related by the constraint
\be
    \Lambda_{1}=\frac{18 k \,\sigma^{2}}{3\sigma^{2}-2}\,.
\ee
In what follows we choose the parameters in the bulk so that both
$k/(3\sigma^{2}-2)$ and $\Lambda_{2}$ are positive, which implies
that $k = 1$. The causal structure is very similar to that of a
S-brane (see Fig.~\ref{caso1}).

Constructing a boundary brane as before, we must satisfy a
junction condition for the dilaton in addition to the metric
condition already discussed.  We
find the scalar condition
\be\label{junscalar}
    4\,
    \,\eta\cdot\partial\phi\,\Big|_{r_{b}=a(\tau)}
               =\frac{\partial}{\partial \phi}\left[
    \rho \,U(\phi) 
          \right]\,\Big|_{r_{b}=a(\tau)}
\ee
where $\eta^\mu= \pm r^{3\sigma^2/2}(\dot r h^{-1}, h\,r^{-3
\sigma^{2}} \dot t, 0,0,0)$ is the unit normal to the brane, and a
dot denotes differentiation with respect to the cosmic time
$\tau$. For a given bulk field configuration, this condition must
be read as a condition on the interaction function $U$.
 This condition can be combined with the Israel junction conditions
to get the energy conservation equation in this case
\be \dot\rho + 3H(\rho + p) = -\rho \,\dot \phi \frac{U'}{U}\,,
\ee
where $' = d/d\phi$. This equation tells us that there is a
nonvanishing energy flux between the bulk and the brane, which
depends on the choice of the scalar coupling $U$.

The junction conditions for the metric give us the brane
trajectory, $r_b(\tau)$, and from this we obtain the effective
Friedmann equation seen by the brane-bound observer. We obtain
\be\label{frietre}
    H^{2}= \frac{\left[ \rho\, U(\phi) \right]^{2}}{144}
    +\left[ \frac{8}{(3 \sigma^{2}-2)(4+3 \sigma^{2})\,a^{2}}-
    \frac{2 \Lambda_{2}}{3(8-3 \sigma^{2}) \, a^{3 \sigma^{2}}}
    \right] \,.
\ee
The first term here gives the dilatonic generalization of the
energy-density contribution to the Friedmann equation. The last
term is the negative-energy contribution, which scales with $a$ as
would a fluid having equation of state parameter $\omega = -1 +
\sigma^{2}$. In order to obtain positive acceleration, we must
choose $\sigma^{2}> \frac23$. A part from this requirement, we can
tune the degree of acceleration of the model by appropriately
dialing $\sigma$. The necessity for the negative-energy
contribution, despite having chosen $k = +1$, can be seen from the
second-last term, which is what would have been expected for $k =
-1$ rather than $k = 1$. This difference is due to the effect of
the $\Lambda_1$ term in the scalar potential, which generates
terms which scale as $1/a^2$ and so which competes with the
spatial-curvature term.

We now return to face the issue of determining the coupling
function $U(\phi)$. For definiteness we choose the ansatz where
$U$ is an exponential: $U(\phi) = e^{-\beta \phi}$. The scalar
junction condition (\ref{junscalar}) can then be rewritten as:
\be
   12\, \frac{\sigma}{a} \sqrt{h \, a^{-3\sigma^2} + \dot a^2} = -\rho \beta
    a^{-3\sigma\beta} \,.
\ee
This condition is equivalent to the Friedmann equation --- and so
is automatically satisfied --- provided we choose $\beta= \pm
\sigma$. Any other value gives a constraint incompatible with the
Friedmann equation, and so one which cannot be true identically
for all values of $\phi$ evaluated at the brane.

Choosing this value, the system satisfies all the junction
conditions, and the Friedmann equation becomes
\be\label{frietreexpl}
    H^{2} =
     \frac{\rho^{2}}{144}\,a^{\mp 6\sigma^{2}}
    + \left[ \frac{8}{(3 \sigma^{2}-2)(4+3 \sigma^{2})}\frac{1}{a^{2}}-
\frac{2 \Lambda_{2}}{3(8-3 \sigma^{2})}\frac{1}{a^{3 \sigma^{2}}} \right] \,,
\ee
while the equation of conservation of energy becomes
\be\label{noncons2}
\dot\rho + 3H(\rho + p) = \pm 3 H \sigma^{2} \rho \,.
\ee
From this equation,  the dependence of the energy density on the scale 
factor, in the case of one form of matter, can be easily found to be:
\be\label{noncons3}
\rho  =  \rho _0 \, a^{-3(1+\omega \mp\sigma^{2})} \,.
\ee
Plugging this into (\ref{frietreexpl}), it is easy to see that the 
unconventional factor $a^{\mp 6\sigma^{2}}$ of the first term
 cancels away. 
That is,  we get simply $H^2 = a^{-6(1+\omega)}/144+$(bulk contributions),
 as  in the nondilatonic 
cases.

The previous two formulae indicate that the resulting cosmology
during the bouncing phase is quite non-standard, due to the flow
of energy from brane to bulk. For this reason, we consider the
model as suitable for describing only the very early universe,
rather than the post-BBN universe which is visible through
cosmological observations. In the following we concentrate on the
Friedmann equation to describe the bouncing properties of this
universe.

\smallskip

Equation (\ref{frietreexpl}) is not easy to solve analytically, so we use
instead the effective potential method to determine if and when a
bounce occurs. The effective potential in this case is given by
\be\label{W1}
    W(a) = \frac{2 \Lambda_{2}}{3(8-3 \sigma^{2})}
\frac{1}{a^{3\sigma^{2}-2}} -
\frac{8}{(3 \sigma^{2}-2)(4+3 \sigma^{2})}-
    \frac{\rho^{2}}{144\, a^{6\sigma^{2}-2}} \,,
\ee
where for definiteness we choose $\beta = -\sigma$. It is also
convenient to choose for simplicity only tension as the matter
content on the brane, $\rho=\lambda=$ constant.
 The dynamics of the bounce changes its dependence on the choice of
the conformal factor $\sigma$. In any case, it is possible to
check that the turning point lies inside the Cauchy horizon (see
Fig. \ref{bo1} for a specific example in which we plot the
potential and the function $h$).

\FIGURE{
\let\picnaturalsize=N
\def\picsize{2.5in}
\def\picfilename{dila1.eps}
\ifx\nopictures Y\else{\ifx\epsfloaded Y\else\input epsf \fi
\let\epsfloaded=Y
\centerline{\ifx\picnaturalsize N\epsfxsize \picsize\fi
\epsfbox{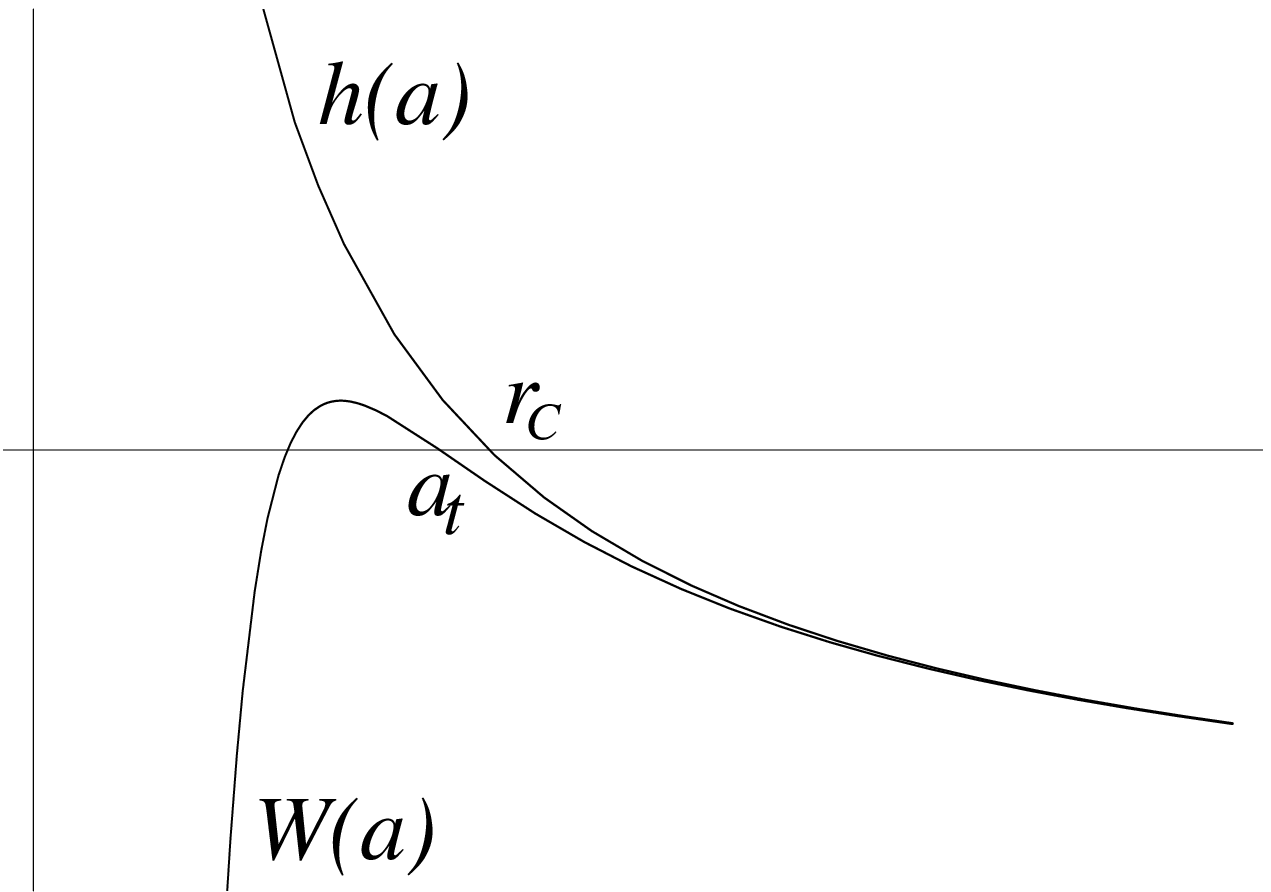}}}\fi
\caption{Effective potential for the dilatonic model described in
eq.~(\ref{W1}). We choose $\sigma=1$, $\Lambda_{2}=33$, $\lambda=3.8$.
The metric
function $h(a)$, given in (\ref{ADSbrane}), is also included. }\label{bo1}}

A full stability analysis has not been performed for these
spacetimes, but because their causal structure is similar to the
pure-gravity $S$-brane considered earlier, it is  likely that
the stability issues are also similar. That is, we expect
stability if we are free to choose only compact-support initial
conditions and to ignore radiation coming out of the
singularities, but not generally otherwise.

\subsection{S0-Branes with a Bulk Dilaton and a Gauge Field}
Our next example is a generalization of the previous one, in such
a way as to change the causal structure of the bulk geometry so
that it does not contain the time-like singularities of the
$S$-brane geometries considered up to this point. The geometry
instead contains naked space-like singularities, as well as the
corresponding Cauchy horizons.

 The model consists of a five-dimensional action, containing
gravity, a dilaton field (with a potential), and a gauge field
which couples non-minimally to the dilaton field (such as
generically occurs in supergravity models). Again we construct a
four-dimensional boundary brane, which in general is also coupled
to the dilaton field. The total action takes the form
\be
    S=\int d^{5} x \sqrt{-g} \left[ R-\frac{1}{2} (\partial \phi)^{2}
    -\frac{1}{4} \, e^{-\sigma\phi}\, F_2^2 - V(\phi)\right]
    +\int d^{4} x \sqrt{-g_{4}}\,\rho\, U(\phi)\,,
\ee
where $F_2 = dA$ is the field strength associated with the gauge
field. We assume a dilaton potential of the form
\be
    V(\phi)= \sum_{i=1}^{3}\,\Lambda_i \,e^{-\lambda_i \, \phi}\,,
\ee
and, as before, $\rho$ is the energy density of any fields which
are localized on the brane (including the brane tension) and
$U(\phi)$ is the dilaton-brane coupling function.

A solution to the bulk field equations for this system is given in
\cite{bnqtz}, and is given by
\bea
    && ds_{5}^{2} = -h(r) d t^{2}+\frac{ d r^{2}}{g(r)} +
    r^{2} d \Omega_{k,3}^{2} \,, \nn \\
    && \phi = \sqrt{3} \,\xi \ln r \,,\nn \\
    && F^{tr} = Q r^{\sigma\xi\,\sqrt{3} -3 -\xi^2/2}\, \epsilon^{t\,r}\,,
\eea
with $g(r) = r^{-\xi^2}\, h(r)$, and
\be
    h(r)= -2M r^{\xi^2/2 - 2} - \frac{2\lambda_1\Lambda_1\,
    r^{\xi^2}}{\xi\sqrt{3}(4+ \xi^2)} -
    \frac{2\Lambda_3\,r^{2}}{3 (8 -\xi^2)} 
    + \frac{[\sigma\,Q^2 - 2\lambda_2\Lambda_2]\,r^{\xi^2 +2
    -\xi\,\lambda_2\,\sqrt{3}}}{\xi\sqrt{3}(8 + \xi^2
    -2\sqrt{3}\,\lambda_2\xi)}\,.
\ee
$\xi$ is an integration constant, and the following constraints
must be satisfied:
\bea
    && \lambda_1 = \frac{\sigma +\lambda_2}{n} \,, \qquad \qquad
    \lambda_3 = \frac{2}{3\lambda_1}\,, 
    \qquad \qquad ( \sigma +\lambda_2) \, \xi = 2\sqrt{3}\,, \nn \\ \\
    &&  \hskip1cm
    \Lambda_1 = \frac{6k\,\lambda_3}{\lambda_3 - \lambda_1}\,, \qquad \qquad
    Q^2 (\sigma + \lambda_3) = 2\Lambda_2(\lambda_2 - \lambda_3)\,.
\eea

\FIGURE{
\let\picnaturalsize=N
\def\picsize{3.5in}
\def\picfilename{SchadSRN.eps}
\ifx\nopictures Y\else{\ifx\epsfloaded Y\else\input epsf \fi
\let\epsfloaded=Y
\centerline{\ifx\picnaturalsize N\epsfxsize \picsize\fi
\epsfbox{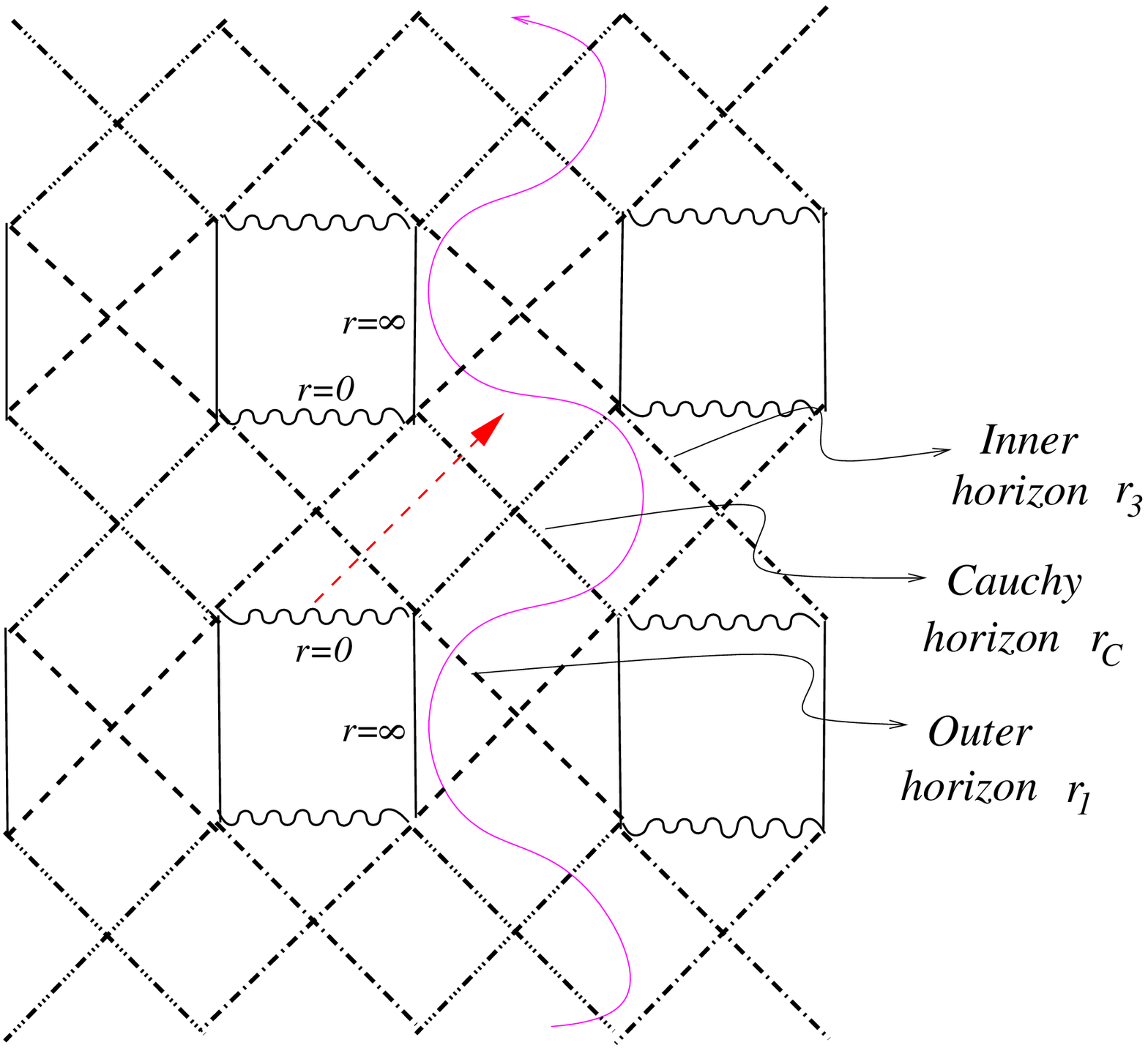}}}\fi
\caption{Penrose diagram for solution (\ref{3horizons}). Notice the
presence of space-like naked singularities, from which signals
can affect brane physics. As a consequence of these naked
singularities,
Cauchy horizons are present.}\label{bo3}}

In order to obtain an example with an interesting bounce, we
specialize to the case where $\xi$, $\Lambda_1$ and $M$ are all
positive, $\lambda_2 = 0$ and $\Lambda_2$ and $\Lambda_3$ are both
negative. Assuming $\lambda_1 = 1/\sqrt{3}$, we find $\xi =2$,
$\lambda_3 =2/\sqrt{3}$, $\sigma= \sqrt{3}$, $\Lambda_1 = +12$
($k=1$) and $Q^2 = -\frac{4}{5} \Lambda_2$. With these values the
bulk solution becomes
\bea\label{3horizons}
    && ds_{5}^{2} = -h(r) d t^{2}+\frac{ r^4}{h(r)}  dr^{2} +
    r^{2} d \Omega_{1,3}^{2} \,, \nn \\
    && \phi = 2\sqrt{3}\, \ln r \,,\nn \\
    && F^{tr} = Q r\,\epsilon^{t\,r}\,, \nn \\
    && h(r)= -2M - \frac{r^4}{2} + \frac{|\Lambda_3|}{6}\,r^{2}
    + \frac{Q^2}{24}\,r^6\,\label{hthis} \, .
\eea
The causal structure of this spacetime is illustrated in the
Penrose diagram of Fig.~\pref{bo3}. The diagram contains
non-standard examples of Cauchy horizons, associated with the
presence of naked  space-like singularities. These singularities
 are naked since signals that follow time-(or null-)like
trajectories, coming from these singularities, can influence
an observer that moves in the geometry.

We now work out the implications of the junction conditions at the
position of the brane\footnote{
Also in this case, the equation of conservation of energy is not
satisfied on the brane alone due to a flow of energy into the
bulk. Considerations similar to the previous model apply.}. The
metric condition is obtained as before, and gives rise to the
following Friedmann equation
\be \frac{\dot
    a^2}{a^2} = \frac{\left[ \rho \, U(\phi) \right]^2 }{144}-
      \frac{h(a)}{a^6}\,,
\ee
where $a(\tau) = r(\tau)$, as before. From the dilaton junction
condition eq.~(\ref{junscalar}), we find
\be
    \frac{8\,\sqrt{3}}{a} \sqrt{h a^{-4} + \dot a^2} = -\rho \beta
    a^{-2\sqrt{3}\beta} \,.
\ee
where the normal vector to the brane is given by $\eta^\mu = \pm
r^{2}(\dot r h^{-1}, \, \dot t h\,r^{-4},0,0,0)$, and we have
taken, as before, an exponential ansatz: $U(\phi) =
e^{-\beta\phi}$.

The scalar junction condition is equivalent to the Friedmann
equation if we choose $\beta= \pm 2/\sqrt{3}$, and with this
choice the system satisfies all the junction conditions. The
effective 4D Friedmann equation for brane-bound observers becomes
\be
    H^{2} = \frac{\rho^2}{144} a^{\mp 8} - 
     \left[-\frac{2M}{a^6} + \frac{|\Lambda_3|}{6\,a^4}- \frac{1}{2\,a^2} 
    + \frac{Q^2}{24} \right]\,.
\ee
Here again, the conservation equation implies the modified dependence 
\be
\rho = \rho _0\, a^{-3(1+\omega \mp 4/3)} \,,.
\ee
Plugging this into the Friedmann-like equation cancels the  extra power of
 the scale factor in the first term, $a^{\mp 8}$. We end up then with 
$H^2 \propto a^{-6(1+\omega)}$ as in the nondilatonic cases. Moreover, in
 the case of a cosmological constant $\omega=-1$, living on the brane, 
 one could cancel it with the charge term in the Friedmann equation, as in
 the Randall-Sundrum-like  scenarios.

 Notice that we do not have to impose additional junction
conditions for the gauge field, since we choose it with opposite
charges at the two sides of the brane. Consequently, the flux
lines extend continuously over the brane, which does not carry any
charge.

We can now analyze if there exist a bouncing brane world, by
looking at the effective potential $W(a)$. Choosing the brane
matter to be pure tension, we find
\be\label{W2}
    W(a)=\, -\frac{2M}{a^{4}}-\frac{1}{2}+\frac{|\Lambda_{3}|}{6 a^{2}}
    +\frac{Q^{2} \, a^{2}}{24} - \frac{\lambda^{2}}{144\,a^{6}}\,.
\ee
The resulting potential is drawn --- together with the metric
function $h$ --- in Fig.(~\ref{bo2}), where we choose the remaining
parameters to lie in regions for which there are three independent
horizons, and for a small value for the brane tension. From this
figure it is clear that there are three turning points, one
corresponding to a short-lived expanding-then-contracting
universe, and the others corresponding to an oscillatory universe
which undergoes repeated expansion and contraction. The innermost
bounce occurs inside the Cauchy horizon (marked as $r_C$ in
figures \ref{bo3} and \ref{bo2}) but the Penrose diagram shows why
this is not inconsistent with having repeated bounces.

\FIGURE{
\let\picnaturalsize=N
\def\picsize{2.5in}
\def\picfilename{dila2.eps}
\ifx\nopictures Y\else{\ifx\epsfloaded Y\else\input epsf \fi
\let\epsfloaded=Y
\centerline{\ifx\picnaturalsize N\epsfxsize \picsize\fi
\epsfbox{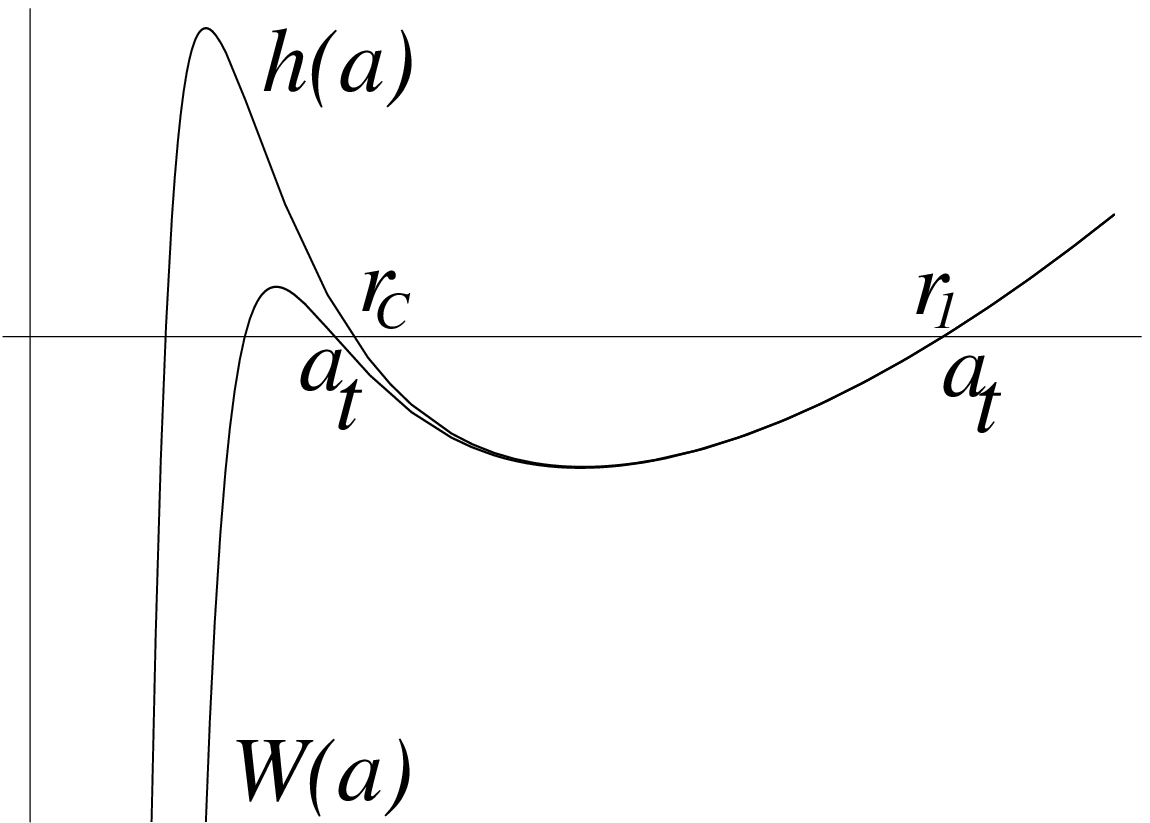}}}\fi
\caption{Effective potential (\ref{W2}), for the charged dilatonic
model. For the parameters, we choose $M=10^{-2}$,
$\Lambda_{3}=Q=1.6$, $\lambda=0.06$. The metric function $h(a)$,
given in (\ref{hthis}), is also included.} \label{bo2}}

The question of the global stability of this non-standard bulk
geometry (and consequently of the brane-world model) remains open,
since it has not yet been addressed in the literature. We expect
the main issues to be similar to what is encountered in the
simpler $S$-brane examples.

\section{Bouncing Probe Branes}
We next turn to a different class of bouncing models which are more closely
based on string theory. In them the brane world is a probe brane
rather than a boundary brane, which we take to move in the
background fields of a stack of source branes, which solve the
field equations of a higher-dimensional supergravity. For the
present purposes such probe-brane systems have both advantages and
disadvantages relative to the codimension-one boundary branes of
the previous section. Their main advantages are that they allow
the discussion of bounces to be moved to a broader context than
purely 5-dimensional examples. In particular, they allow the use
of supersymmetric backgrounds for which the bulk is known to be
stable. Their main disadvantage is that the observer's brane is
just a probe, and so the analysis does not allow one to follow how
the brane stress-energy back-reacts onto the bulk spacetime.
 Other examples of probe brane cosmologies in  various string theory 
backgrounds have  been discussed recently in \cite{kiritsis,kachru,brax}.

By taking a background that preserves some of the supersymmetries
of the bulk action we can be sure the background geometry is
stable. This leaves only instabilities associated with the branes
themselves as potential threats to the stability of the bouncing
brane-world. When the back-reaction of the probe branes is taken
into account new instabilities can appear, but these types of
instabilities can be controlled if the system is close to a
supersymmetric limit. One easy example of this type consists of a
probe $D$-brane moving in the fields of a stack of parallel
$D$-branes having the same dimension. When the probe brane is at
rest this system is supersymmetric, but small relative velocities
break the supersymmetry very softly, allowing a time-dependent
metric to appear on the probe brane. Within the non-relativistic
approximation there is no interaction potential, and the probe
brane trajectories can be made arbitrarily close to straight
lines. In such a situation there is a bounce in the probe-brane
induced metric as the branes pass through their point of closest
approach to one another.

\subsection{The Branonium System}
\FIGURE{
\let\picnaturalsize=N
\def\picsize{2.5in}
\def\picfilename{branonium.eps}
\ifx\nopictures Y\else{\ifx\epsfloaded Y\else\input epsf \fi
\let\epsfloaded=Y
\centerline{\ifx\picnaturalsize N\epsfxsize \picsize\fi
\epsfbox{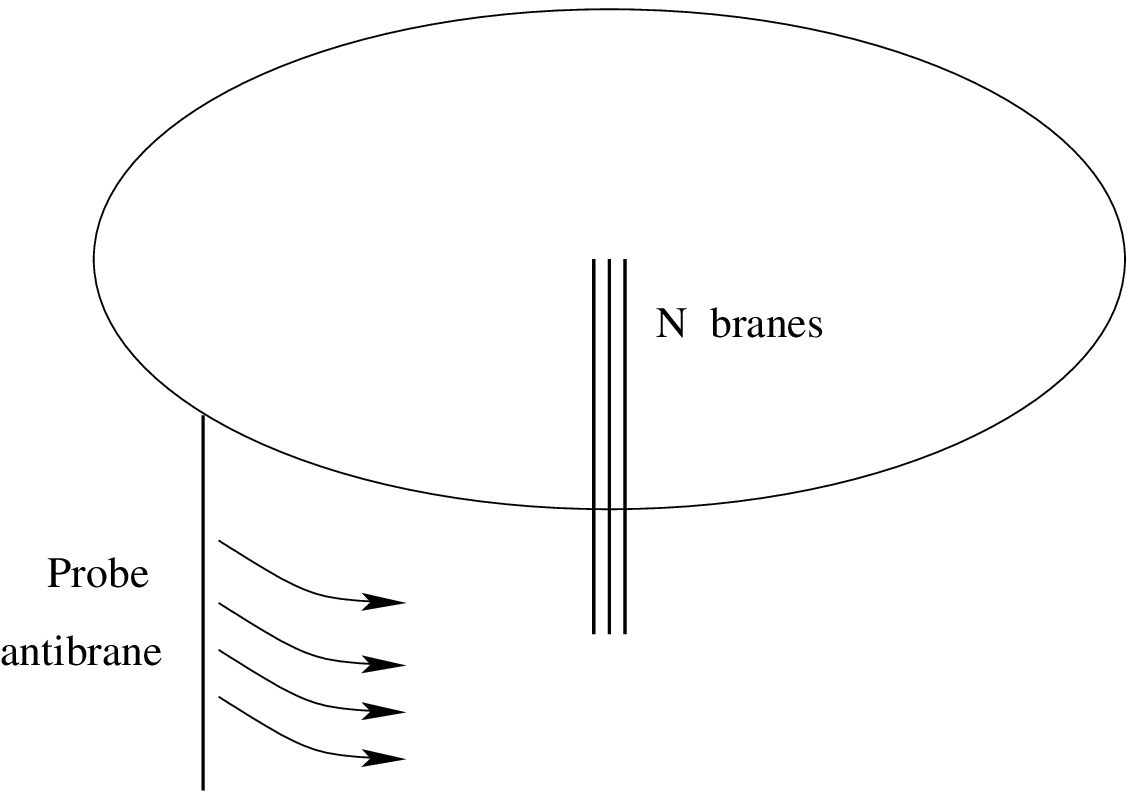}}}\fi
\caption{\small Branonium is the bound state of an anti-brane
orbiting a set of source branes.} \label{branonium}}
%
%
We briefly review here the analysis of \cite{branonium1}, who
consider the motion of a straight probe anti-brane as it moves
within the fields set up by a stack of $N$ parallel source branes
(see figure \ref{branonium}). The source and probe branes are
imagined to be parallel and to have the same dimension, $d = p+1$.
It is argued in \cite{branonium1} that a probe brane which is
initially parallel to the source branes --- and which starts
sufficiently far away --- tends to move rigidly, without bending
or rotating relative to the source branes. Because of this the
probe-brane dynamics is described by the motion of its center of
mass, which behaves much as would a point particle moving through
the dimensions transverse to the branes.

Since the branes are flat their internal dimensions are easily
compactified using toroidal compactifications, and although more
care is required the same can also be done for the dimensions
transverse to the brane. We imagine that once this is done there
are $d_T$ dimensions transverse to the brane which are relatively
large and within which the probe brane moves.

\subsubsection*{Dynamics and Orbits}
Following \cite{branonium1} we take the bulk fields to be governed
by the bosonic supergravity action
\begin{equation}
    \label{eq:EFbulkaction}
    S_s = - \int d^Dx \, \sqrt{- g} \left[ \frac12 \, g^{MN} \Bigl(
    R_{MN} + \partial_M \, \phi \, \partial_N \, \phi \Bigr) + {1
    \over 2 \, n!} \;e^{\alpha \phi} \, F_{M_1...M_n} \, F^{M_1...M_n}
    \right]\,,
\end{equation}
where the $n$-form field strength is related to its $(n-1)$-form
gauge potential in the usual way $F_{[n]} = dA_{[n-1]}$. $n$ is
related to the spacetime dimension $d$ of the source branes by $d
= n - 1$.

The solution describing the fields sourced by a stack of source
branes is given in the Einstein frame by
\be \label{Dbrane}
    ds^2 = h^{-\tgamma} dx^2 +  h^{\gamma} dy^2 \, ,\qquad e^{\phi} =
    h^{\kappa} \, ,\qquad A_{01...p} = (1 -  h^{-1}) \, ,
\ee
where all the other components of the $(n-1)$-form field vanish.
The constants $\gamma$, $\tgamma$ and $\kappa$ are given by
\be \label{braneconsts}
    \gamma = {d \over (D-2)} \, , \qquad
    \tgamma = {\td \over (D-2)} \, , \qquad
    \kappa = {\alpha \over 2} \, ,
\ee
where $\td = D - d - 2$ and for $d = p+1$ dimensional branes
$\alpha = \frac12(3-p)$. We denote the $d$ coordinates parallel to
the branes by $x^\mu$ and the $d_T = D-d$ transverse coordinates
by $y^m$. The harmonic function, $h$, is given by $h = 1 +
k/r^\td$ with $k > 0$ and $r^2 = y^m y_m$.

The Lagrangian describing the probe brane motion through these
fields is obtained by evaluating the Dirac-Born-Infeld (DBI) and
Wess-Zumino (WZ) action at the position of the probe brane, using
the fields
--- dilaton, $\phi$, metric $g_{MN}$ and Ramond-Ramond $d$-form
gauge potential $A_{M_1...M_d}$ --- sourced by the stack of source
branes. For some choices of parameters this leads to the simple
lagrangian for the motion of the probe brane's center of mass
\begin{equation} \label{ParticleL}
   {\mathcal  L} = -  \frac{m}{h} \left( \sqrt{1 - h \, v^2} -
    {{q}} \right) \, .
\end{equation}
Here $v^2 = \dot{y}^m \dot{y}_m$ denotes the squared speed of the
probe brane's center of mass. The mass, $m$, is related to the
brane tension, $\cT_p$ and spatial world volume, $V_p$, by $m =
\cT_p V_p$, and we imagine the dimensions parallel to the branes
to be compactified so that $V_p$ is very large, but finite.
The first term of eq.~\pref{ParticleL} represents the
probe-brane coupling to the dilaton and metric through the DBI
action, while the second term gives its coupling to Ramond-Ramond
potential through the WZ term. The probe brane's charge for this
gauge potential is denoted ${q}$, and is $+1$ for a probe brane or
$-1$ for an antibrane.

Conservation of angular momentum in the transverse dimensions
ensures that the brane motion is confined to the plane spanned by
the particle's initial position and momentum vectors. Denoting
polar coordinates on this plane by $r$ and $\theta$, and
specializing to D$p$-branes in 10 dimensions (or their
dimensionally-reduced counterparts) the Lagrangian for the
resulting motion becomes
\begin{equation} \label{grangiano}
{\mathcal L} = - \, \frac{m}{h} \left[ \sqrt{1 - h \, (\dot{r}^2 + r^2
\dot{\theta}^2)} - q \right] \, ,
\end{equation}

Remarkably, the orbits for this fully relativistic branonium
Lagrangian can be found by quadrature \cite{branonium1} simply by
following the standard steps used for nonrelativistic
central-force problems, and are given by:
\begin{equation}
    \theta - \theta_0 = \int_{1/r_{0}}^{1/r}  \frac{dx}{\sqrt{A + B \,
    x^\td - x^2}} \label{teta}
\end{equation}
where $A = (\varepsilon^2+ 2 \, q \, \varepsilon)/\ell^2$ and $B =
k \, \varepsilon^2/\ell^2$. Here $\varepsilon = E/m$ is the energy
per unit mass and $\ell = p_\theta/m$ is the angular momentum per
unit mass, which are given explicitly by
\begin{equation}
    \ell = \frac{p_{\theta}}{m}  = \frac{r^2 \dot{\theta}}{
     \sqrt{1 - h \, (\dot{r}^2 + r^2 \dot{\theta}^2)}}
\end{equation}
and
\begin{equation} \label{energia3}
    \varepsilon = \frac{E}{m} = \frac{1}{h}
    \left[ 1 + h \,
    \left( \frac{\ell^2}{r^2} + \rho_r^2 \right)
    \right]^{1/2} - \frac{{q}}{h}
    \, ,
\end{equation}
both of which are conserved during the motion (up to
Hubble-damping effects). Here $\rho_r$ is the canonical momentum
in the radial direction, given by
\begin{eqnarray}
    \rho_r = \frac{p_{r}}{m}  &=&  \frac{\dot{r}}{
    \sqrt{1 - h \, (\dot{r}^2 + r^2 \dot{\theta}^2)}} \, ,
    \nonumber \\
    &=& \dot{r} \; \left( \frac{1 +
    \ell^2 \, h /r^2}{1 - h\, \dot{r}^2}
    \right)^{1/2} \, .
\end{eqnarray}

The turning points of the motion may be found by examining the
effective potential, which is obtained by evaluating the energy at
$\rho_r = 0$, since this is an absolute lower bound for the
energy. The result is
\be \label{effpot}
    V_{\rm eff}(r) =
    \frac{1}{h} \left\{ \left[ 1 + h \,
    \left( \frac{\ell^2}{r^2} \right) \right]^{1/2}
    - {{q}} \right\} \, ,
\ee
where $\ell$ is the orbital angular momentum. We plot this
potential for the choice $\td = 1$ (3 large transverse dimensions)
in Fig.~\pref{VeffFig}. Classically allowed motion occurs for any
energy, $\varepsilon$, which lies above this curve, and turning
points occur for $r= r_t$ satisfying $\varepsilon = V_{\rm
eff}(r_t)$. Clearly if $\ell \ne 0$ at least one turning point
exists for any energy, corresponding to the point of closest
approach due to the centrifugal barrier of a probe brane having a
nonzero initial impact parameter. A second turning point occurs
when bound orbits exist, such as happens for the brane-antibrane
example ($q = -1$) with $\td = 1$.

\FIGURE{
\let\picnaturalsize=N
\def\picsize{3.0in}
\def\picfilename{potential.eps}
\ifx\nopictures Y\else{\ifx\epsfloaded Y\else\input epsf \fi
\let\epsfloaded=Y
\centerline{\ifx\picnaturalsize N\epsfxsize \picsize\fi
\epsfbox{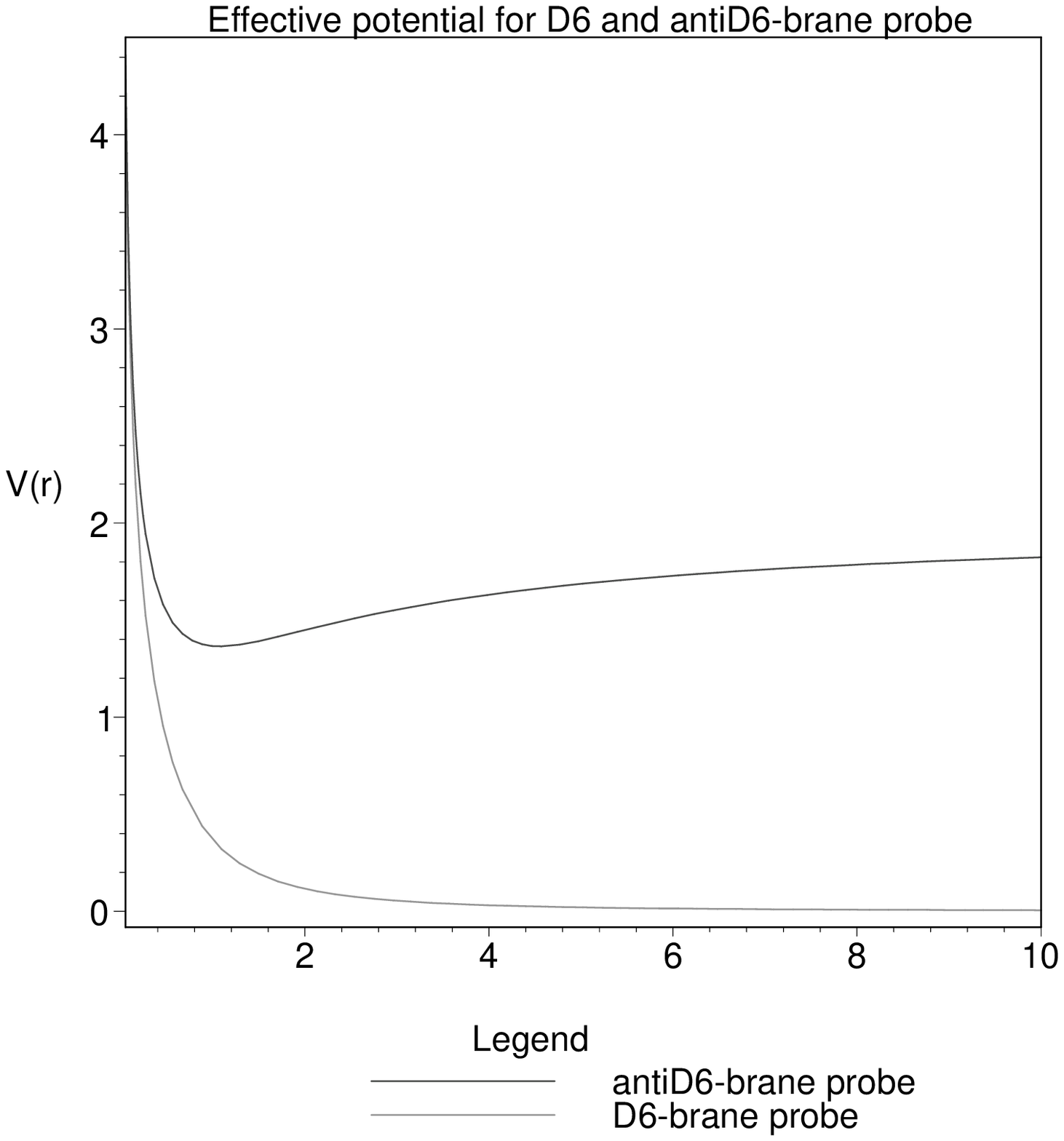}}}\fi
\caption{\small The effective potential for the radial motion of
the brane centre-of-mass.} \label{VeffFig}}

\subsection{Branonium Bounces}
Degrees of freedom trapped on the probe brane `feel' the following
induced metric
\begin{equation}\label{brametr}
    d\hat{s}^2 =  h^{-\beta} \left[ \left(
    - 1 + h v^2 \right) dt^2 + (d\xi^{i})^2\right] \,,
\end{equation}
where $i=1,\dots,p$, and
\bea
\beta =  \left\{
\begin{array}{ll}
\frac{8-d}{8} &   {\rm  in \,\,\,the \,\,\, Einstein\,\,\, frame}\\ \\
\frac{1}{2}
 \; \; \; \; & {\rm in \,\,\,the \,\,\, string\,\,\, frame }
\end{array}
\right.
\eea

To go to the string frame one must multiply the Einstein-frame
metric by $e^{\lambda \phi}$, where $\lambda=1/2$. Notice that for
the (interesting) case $d=4$ the value of $\beta$ in the two
frames coincide.
 The metric (\ref{brametr}) is explicitly time-dependent, despite
the static nature of the bulk geometry, due to the brane's motion
through the bulk. The time dependence appears through the
dependence of the harmonic function, $h = 1 + k/r^\td$, on the
probe brane position as well as through the explicit dependence on
the brane speed if the brane accelerates.

The induced metric on the brane has the FRW form, with flat
spatial slices, $k = 0$, as may be seen by transforming to the FRW
time coordinate, $\tau$, defined by
\begin{equation}
    \frac{d\tau}{dt} = \sqrt{h^{-\beta} (1 - h v^2)} \, ,
\end{equation}
in terms of which the metric takes the form
\begin{equation}
    ds^2 = - d\tau^2 + a^2(\tau) (d\xi^{i})^2
\end{equation}
with scale factor $a(\tau) = h^{-\beta/2}$.

Since $\beta > 0$ the scale factor increases when $h$ decreases,
and so when the distance between the branes increases. For
circular orbits there is no time-dependence to the scale factor at
all, but this is no longer true for elliptic orbits for which the
scale factor oscillates, making distances on the brane appear to
contract when the branes approach one another and expand when they
recede. Clearly we expect bounces to occur since classical
trajectories exist for which the radial motion is not monotonic.

The Hubble parameter for this geometry has the form
\begin{equation}
    H = \frac{1}{a}\frac{da}{d\tau} =
    \frac{1}{a}\frac{da}{dt}\frac{dt}{d\tau} = \frac{\td  \beta  k}{2
    \, r^{\td+1}} \frac{h^{\beta /2 -1}}{\sqrt{1 - h v^2}} \,
    \frac{dr}{dt} \, .
\label{hubblebran}
\end{equation}
Notice that the sign of $H$ depends only on the sign of the radial
velocity, as advertised. $dr/dt$ can be eliminated in favor of the
energy $\varepsilon$ and angular momentum $\ell$ using
\be
    \frac{\dot{r}^2}{1 - h \, v^2} =
    \rho_r^2 = \frac{(\varepsilon \, h + q)^2 - 1}{h} -
    \frac{\ell^2}{r^2} \,,
\ee
to write the effective 4D Friedmann equation (using $q^2 = 1$)
\begin{equation}
    H^2 = \left( \frac{\td  \beta  k}{2 \, r^{\td+1}} \right)^2 h^{(\beta -2)
    }\,  \left[ \varepsilon (\varepsilon \, h + 2\, q)  -
    \frac{\ell^2}{r^2} \right]\, .
\end{equation}
As expected, for probe branes ($q = +1$) it is the angular
momentum term which plays the role of the negative contribution
which is responsible for the bounce. Notice that there is no
$\rho_r$ dependence in the Friedmann equation because our
probe-brane assumption implies the energy on the brane is so small
that its back reaction on the metric is negligible. The scaling
with $a$ of the various terms on the right-hand-side takes a
complicated form, obtained by inverting $a = h^{-\beta/2}(r)$ to
obtain $r(a)$.

A bounce is present when the right-hand side of the previous
formula vanishes for a finite value of $r$ (and, consequently, of
$a$). For the particular and interesting case $p=3$ (corresponding
to a four dimensional cosmology) one finds a solvable equation,
that furnishes the following two turning points, $r_{12}$
\be
    r_{12}^{2}\,=\,\frac{1}{2\varepsilon(\varepsilon+2q)}\left[
    \ell^{2}\pm\sqrt{\ell^{4}-4 k \varepsilon^{3}(\varepsilon+2q)} \right]\,.
\ee
There are two turning points, and so we have a cyclic model, when
the following inequalities are satisfied
\bea
    &\varepsilon& +2 q \,>\,0\,\nonumber \\
&\ell^{4}& \, >\, 4\varepsilon^{3} k (\varepsilon+2q)\,. \eea

Also notice that in the non-relativistic limit we have $\ell^2/r^2
\ll 1$, $k/r^\td \ll 1$ and $\varepsilon \approx 1 - q + \delta
\varepsilon$ with $|\delta\varepsilon| \ll 1$, and so to leading
order in $k/r^\td$ and $v^2$ we may take $h \to 1$ in the above
expression for $H^2$ (because of the overall pre-factor of
$(k/r^{\td+1})^2$). This gives the approximate result
\begin{equation}
    H^2 \approx \left( \frac{\td  \beta  k}{2 \, r^{\td+1}} \right)^2
    \,  \left[ 2 \delta \varepsilon \left[ (1-q) h + 2q \right] + 2 (1-q) (h-1)-
    \frac{\ell^2}{r^2} \right]\, .
\end{equation}
For branes ($q = 1$) the potential vanishes and the only thing
that remains is the angular momentum term and the variation of the
energy, consistent with the supersymmetry of the brane-brane
system, which has no interaction at leading order. Trajectories
are, at this level, straight lines. The Hubble constant becomes
zero when $r$ has the smallest value, which is the same as the
initial impact parameter. For antibranes ($q = -1$) we have a
potential $2 (1-q) k/r^\td$, which is attractive as expected.
Notice that the term in the brackets vanish at $4k/r^\td
-\ell^2/r^2 = 0$

\subsection{Compactification}
As before, it is important to ask in the present instance
whether or not the probe brane of the previous discussion
experiences a bounce in the 4D Einstein frame. For systems of
parallel branes and antibranes it is relatively straightforward to
compactify the dimensions along the brane dimensions, because
translation invariance in these directions allows this to be done
on a torus. Such a compactification does not alter the discussion
of the orbits given above.

Because the branes involved are BPS, toroidal compactification of
the dimensions transverse to the branes is also possible. This is
because the BPS condition allows a general solution for the bulk
supergravity fields to be found for arbitrary collections of
parallel and static source branes, and toroidal compactification
may be achieved by supplementing the original source branes with
appropriate images. If the transverse dimensions are large
compared with size of the probe brane orbits, the image charges
should not appreciably perturb the motion, leaving the above
analysis unchanged (just taking into account the cancellation of Ramond-Ramond charges in the compact space). 

If we start, as we have assumed until now, in the
higher-dimensional Einstein frame, any dimensional reduction
introduces a factor of the volume of the compact dimensions into
the effective 4D Planck mass, but this volume is independent of
time for our purposes because the bulk fields are static and the
volume does not depend on the probe-brane position. It follows
that the bounce we describe in this section also occurs in the 4D
Einstein frame.

Notice that, however, an observer on the brane that measures the Newtonian
law between two bodies discovers that the force depends on time.
This is due to the fact that the brane, moving through the bulk,
probes different values of the warp factor   at each position of the brane.
This fact reflects on a time dependence of the mass of the bodies, and
this implies a time dependence on the gravitational force~\cite{kachru}.

\subsection{Stability}
As mentioned earlier, stability issues should be easier to
understand in the branonium setup, given the stability of the bulk
field configurations. Although the brane-antibrane system is
certainly unstable to mutual annihilation, this is only effective
once the branes pass to within of order the string length of one
another, $r \lsim l_s$. Ref.~\cite{branonium1} investigated other
stability issues, including the possibility of the probe brane
rotating or bending as it moves in the background fields, and
found that they were stable.

In particular, stability against probe-brane bending in response
to tidal forces from the source branes was found not to be a
problem for branes separated by distances much larger than the
string scale because the brane tension acted as a restoring force
which overwhelmed the disrupting tidal forces. One might worry
that to the extent that the brane is at a classical turning point,
its effective tension vanishes as its classical kinetic energy
does.\footnote{We thank Rob Myers for conversations on this
point.}

We do not believe such a tidal instability to occur because the
above analysis shows that for branonium the bounce is tied to a
vanishing radial velocity, and this is independent of the tension.
In particular, we have seen that bounces are possible in the
nonrelativistic limit, and in this limit the tension becomes a
large additive constant in the probe brane energy, which is
completely independent of the radial motion.

\section{Conclusions}
In this paper we construct several new brane-world models for
which brane observers experience a bouncing cosmology. We present
two classes of examples: those having a boundary 3-brane in a
curved 5-dimensional spacetime, and those involving probe branes
in potentially more than one higher dimension.

Our boundary-brane models came in several variants, depending on
whether the bulk fields consisted of pure gravity, dilaton
gravity, or dilaton-gravity plus a gauge field. In all cases we
used explicit solutions to the bulk equations, and built the
boundary brane by cutting and pasting along the brane's position
in the usual way. The various junction conditions were implemented
and determined the brane's trajectory through the bulk space. In
all cases where the brane geometry bounced, induced bulk-curvature
effects provided the negative-energy contributions to the
effective 4D Friedmann equation which are required on general
grounds.

The bulk geometries obtained had horizons and singularities, which
played an important role in achieving the brane-world bounce, by
providing the required negative-energy terms in the effective 4D
Friedmann equation. We have consequently a higher dimensional,
geometrical picture of the source of DEC violating terms, the
presence of the source singularities for the bulk fields produce
the necessary acceleration.\footnote{The fact that the negative
tension of time-like naked singularities produces acceleration has
been already pointed out in \cite{cck,bqrtz}.}
  For the simplest
geometries these singularities were time-like, but for more
complicated examples they were space-like.

The presence of singularities in the bulk is worrisome from the
point of view of stability since it signals the lack of full
control of the system. In particular, there can be uncontrollable
signals coming from the singularity which could crucially affect
the physics on the observer's brane. This is likely related to the
violation of the dominant energy condition (DEC) which the 4D
observer sees, since this energy condition is used in the proof
that energy and momentum cannot appear acausally, from outside the
observer's light cone.

Although we show that scalar-field perturbations on the 3-brane
are not unstable, this does not preclude the existence of
instabilities in the bulk theory, such as has been considered for
some of the spacetimes considered in \cite{bqrtz} and
\cite{costa}. In the examples which have been examined the
existence of intrinsic bulk instabilities appears to be tied to
assumptions about initial conditions, and to the properties of the
timelike singularities. According to \cite{costa} these geometries
could also be free of bulk instabilities.
  A similar study of the stability of the
models having spacelike singularities has not been done, and we
believe would be worthwhile. If the bulk theory is unstable, it
undermines the use of these models as constructions of brane-world
bounces without instabilities.

Our second class of models consisted of probe branes moving
through the supergravity field configuration set up by a stack of
source branes.  Bounces occur for observers riding on branes which
move through these geometries, since the induced scale factor
depends monotonically on the brane's radial position. Bounces
therefore occur for any classical trajectory that changes its
radial direction.

Stability is under better control in these latter models, because
the bulk space is supersymmetric and so is stable. We did not find
any further instabilities associated with the brane motion, and to
the extent that these are really absent they provide examples of
bouncing brane-world cosmologies within a completely stable
extra-dimensional theory. 
 We expect that a similar behaviour will
occur for the more general D$p$-D$p$' systems discussed in \cite{bgqr}.

 We believe our models present interesting examples where  the smooth
bouncing from contracting to expanding universes could occur. The
problems found in \cite{myers} for previous proposals do not directly
apply to ours and it is an interesting challenge to establish
whether or not   these
are fully stable bouncing universes in the 4D Einstein frame. 

\section{Acknowledgments}
We would like to acknowledge interesting discussions with Nemanja Kaloper, Rob
Myers and Eric Poisson, as well as partial research funding from
NSERC (Canada), FCAR (Qu\'ebec) and McGill University. C.B. and
F.Q. thank the KITP in Santa Barbara for their hospitality while
this work was being completed (as such, this research was
supported in part by the National Science Foundation under Grant
No. PHY99-07949). G.~T.~is supported by the European TMR Networks
HPRN-CT-2000-00131, HPRN-CT-2000-00148 and HPRN-CT-2000-00152.
I.~Z.~was partially supported by CONACyT, Mexico and by the
United States Department of Energy under grant DE-FG02-91-ER-40672.

\end{document}